\documentclass[amsfonts, amssymb, amsmath, aps, prl, reprint, superscriptaddress]{revtex4-1}
\usepackage[english]{babel}
\usepackage[utf8]{inputenc}
\usepackage[dvipsnames]{xcolor}
\usepackage[colorinlistoftodos, color=green!40, prependcaption]{todonotes}
\usepackage{amsthm}
\usepackage{mathtools}
\usepackage{physics}
\usepackage{xcolor}
\usepackage{graphicx}
\usepackage{adjustbox}
\usepackage{placeins}
\usepackage[T1]{fontenc}
\usepackage{lipsum}
\usepackage{csquotes}
\usepackage[pdftex, pdftitle={Article}, pdfauthor={Author}]{hyperref} 
\graphicspath{{figures/}}

\begin{document}
\title{Smooth triaxial weaving with naturally curved ribbons}

\author{Changyeob Baek}
\affiliation{Department of Mechanical Engineering, Massachusetts Institute of Technology, Cambridge, USA}
\affiliation{Flexible Structures Laboratory, Institute of Mechanical Engineering, École Polytechnique Fédérale de Lausanne (EPFL), Lausanne, Switzerland}

\author{Alison G. Martin}
\affiliation{Independent Artist, Fivizzano, Italy}

\author{Samuel Poincloux}
\affiliation{Flexible Structures Laboratory, Institute of Mechanical Engineering, École Polytechnique Fédérale de Lausanne (EPFL), Lausanne, Switzerland}

\author{Tian Chen}
\affiliation{Flexible Structures Laboratory, Institute of Mechanical Engineering, École Polytechnique Fédérale de Lausanne (EPFL), Lausanne, Switzerland}
\affiliation{Computer Graphics and Geometry Laboratory, School of Computer and Communication Sciences, École Polytechnique Fédérale de Lausanne (EPFL), Lausanne, Switzerland}

\author{Pedro M. Reis}
\email[Correspondence email address:]{pedro.reis@epfl.ch}
\affiliation{Flexible Structures Laboratory, Institute of Mechanical Engineering, École Polytechnique Fédérale de Lausanne (EPFL), Lausanne, Switzerland}


\begin{abstract}
Triaxial weaving is a handicraft technique that has long been used to create curved structures using initially straight and flat ribbons. Weavers typically introduce discrete topological defects to produce nonzero Gaussian curvature, albeit with faceted surfaces. We demonstrate that, by tuning the in-plane curvature of the ribbons, the integrated Gaussian curvature of the weave can be varied continuously, which is not feasible using traditional techniques. Further, we reveal that the shape of the physical unit cells is dictated solely by the in-plane geometry of the ribbons, not elasticity. Finally, we leverage the geometry-driven nature of triaxial weaving to design a set of ribbon profiles to weave smooth spherical, ellipsoidal, and toroidal structures.
\end{abstract}


\maketitle

Traditional basketmakers have long been employing the handicraft technique of triaxial weaving to fabricate intricate shell-like structures by interweaving initially straight ribbons into tri-directional arrays~\cite{martin2015basketmaker, ayres2018beyond}. Beyond basketry, triaxial weaving is also encountered in textiles~\cite{dow1970preliminary}, composite materials~\cite{phani2008elastic}, molecular chemistry~\cite{lewandowska2017triaxial, liu2018geometry}, and biology~\cite{brodsky2001biological}. While weaving with straight ribbons in a regular hexagonal pattern yields a flat surface, topological defects (\textit{e.g.}, pentagons or heptagons) induce local out-of-plane geometry~\cite{sadoc2006geometrical, richeson2012euler, martin2015basketmaker, callens2018flat}. Basketmakers have extensive empirical know-how on how and where to place these defects, and recent research has investigated their optimal placements to approximate target surfaces~\cite{ayres2018beyond,vekhter2019weaving}. The strategy to achieve shape by defects is also akin to the concept of topological charge~\cite{nelson1983order} in curved two-dimensional (2D) crystals such as the buckminsterfullerene~\cite{kroto1985c60}, colloidal crystals~\cite{dinsmore2002colloidosomes, bausch2003grain, irvine2010pleats}, confined elastic membranes~\cite{seung1988defects, grason2013universal}, and dimples on curved elastic bilayers~\cite{brojan2015wrinkling, jimenez2016curvature}. However, the curvature attained from these defects is discrete, which limits the range of realizable shapes. Even if previous studies~\cite{mallos2009weave, akleman2009cyclic} have suggested a polygon-based combinatorial design procedure that includes weaving with initially curved ribbons, a predictive understanding of the effect of the ribbon geometry on the shape of the weave is lacking.

\begin{figure}[b]
    \centering
    \includegraphics[width=\columnwidth]{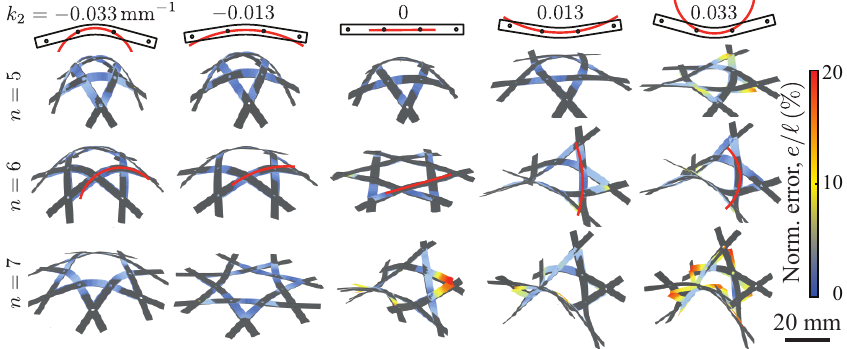}
    \caption{Representative family of triaxially woven unit cells for different numbers of ribbons (rows, $n=\{5,\,6,\,7\}$) and for  different values of the in-plane curvature in their middle segment (columns, $k_2=\{-0.033,-0.013,0,0.013,0.033\}\,\mathrm{mm}^{-1}$). 
    All ribbons have three segments, each of arc length $\ell_1=\ell_2=\ell_3=15\,\mathrm{mm}$; only the middle one is curved ($k_2\neq 0\,\mathrm{mm}^{-1}$), while those in the two extremities are naturally flat ($k_1=k_3=0\,\mathrm{mm}^{-1}$). Tomographic $\mu$CT scans (grey images) are juxtaposed on FEM simulations, color-coded by the distance between their respective centerlines locations, $e/\ell$. See the associated video in~\cite[]{SupInf_PRL}.} 
    \label{fig:1}
\end{figure}

Here, we investigate how triaxial weaving with naturally curved (in-plane) ribbons can yield smooth three-dimensional (3D) shapes. We make use of a combination of rapid prototyping, X-ray micro-computed tomography ($\mu$CT), and finite element methods (FEM) to perform a detailed characterization of the geometry of our woven structures. First, we take a unit-cell approach to systematically explore how the original 2D geometry of the ribbons dictates the 3D shape of the weaves and regard these cells as building blocks to construct more complex woven objects. Fig.~\ref{fig:1} and the associated video~\cite[]{SupInf_PRL} show representative unit cells with different topological characteristics and with ribbons with different in-plane curvatures. Excellent agreement is found between the experiments ($\mu$CT) and simulations (FEM). These unit cells comprise \textit{n} identical ribbons that are woven to form an \textit{n}-gon surrounded by a total of \textit{n} triangles. Each ribbon has three segments (indexed by $j=\{1,2,3\}$), with rivets placed at the crossing points to fix the segment length $\ell_j$. The in-plane curvature, $k_j$, of each segment can be varied continuously (see red solid lines in Fig.~\ref{fig:1}). Traditional weaving corresponds to the case of straight ribbons, $k_j=0\,\mathrm{mm}^{-1}$ (Fig.~\ref{fig:1}, middle column). By considering ribbons that are naturally curved in-plane (examples in Fig.~\ref{fig:1} with $k_2 \neq 0$), we demonstrated that the curvature of the resulting surface of the unit cells can be tuned smoothly, in a way not possible through the traditional approach. A purely geometric analysis is performed to rationalize the integrated Gaussian curvature of the physical unit cells, revealing that geometry is at the core of setting the shape of our physical triaxial weaves. This geometric reasoning forms the basis of a set of design principles, which are then leveraged to construct a variety of smooth canonical structures, including spherical, ellipsoidal, and toroidal weaves. 

\begin{figure}
    \centering
    \includegraphics[width=\columnwidth]{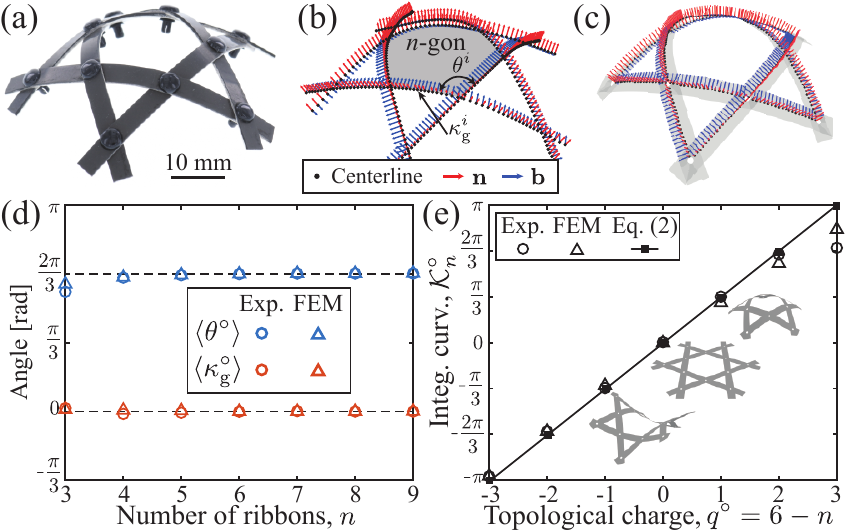}
    \caption{Unit cells woven with straight ribbons. (a)~Photograph of a unit cell with $n=5$ straight ribbons. (b)~Experimental data of the framed centerlines of the cell in (a) extracted from $\mu$CT~\cite[S1.1-3]{SupInf_PRL}. (c)~FEM-computed version of (b). (d)~Average interior angles of the \textit{n}-gon, $\langle \theta^\circ \rangle$, and average integrated geodesic curvatures, $\langle \kappa^\circ_\text{g} \rangle$, \textit{vs.} \textit{n}. The horizontal dashed lines at $\langle \theta^\circ \rangle = 2\pi/3$ and $\langle \kappa^\circ_\text{g} \rangle = 0$ are drawn to aid visual comparison with the geometric prediction.} (e)~Integrated Gaussian curvature of the unit cells with straight ribbons, $\mathcal{K}^\circ_n$, computed through~Eq.~\eqref{eq:gb}, versus $q^\circ=6-n$. The solid line is the prediction from~Eq.~\eqref{eq:topologicalcharge}. Inset: FEM-computed unit cells with $q^\circ=\{-3,-2,\cdots,3\}$, or equivalently $n=\{9,8,\cdots,3\}$. 
    \label{fig:2}
\end{figure}

Before turning to the general case of curved ribbons, we first focus on the `\textit{traditional weaving}' of unit cells with naturally \textit{straight} ribbons ($k^\circ_1=k^\circ_2=k^\circ_3=0$); hereon, the superscript $(\cdot)^\circ$ denotes quantities pertaining to straight ribbons. In Fig.~\ref{fig:2}(a), we present the photograph of a physical unit cell for a representative case with $n=5$ ribbons. 
The specimens were fabricated by, first, laser-cutting ribbons of width $4\,\mathrm{mm}$ from a polymer sheet and, then, hand-weaving them to produce a 3D structure, which was imaged tomographically using a $\mu$CT scanner ($\mu$CT100, Scanco Medical; $a=29.3\,\mu\text{m}$ voxel size). The original polymer sheet was a bilayer of a polyethylene terephthalate (PETE; Young's modulus $E_1\approx\,3\,\text{GPa}$) plate of thickness $t_1=0.25\,\mathrm{mm}$, coated with an elastomer-metal composite ($E_2\approx\,1$MPa) of thickness $t_2=0.35\,\mathrm{mm}$. The latter comprised vinylpolysiloxane (VPS-16, Zhermack) infused with a metal powder (NdFeB, 30065-089, neo Magnequench; $\approx5\,\mu$m particle size) mixed at 2-to-1 weight ratio. Given the disparity in bending stiffnesses of the two layers, $E_1(t_1)^3/[E_2(t_2)^3]\approx \mathcal{O}(10^3)$, the mechanical stiffness of the ribbons was provided by the PETE, with a width-to-thickness ratio of 16. The radiopacity of the elastomer-metal served in detecting of the ribbons using X-ray tomography to extract their framed centerlines from the volumetric data~\cite[S1]{SupInf_PRL} (Fig.~\ref{fig:2}b). The corresponding framed centerlines extracted from FEM  (see~\cite[S2.1]{SupInf_PRL} for procedure) are in excellent agreement with the experiments, as demonstrated in Fig.~\ref{fig:2}(c).

We characterize the shape of the unit cells by quantifying their curvature. However, since the \textit{n}-gon in the woven unit cell does not possess a well-defined surface (its inner region is void of material), it is impossible to define a pointwise Gaussian curvature, $K$. However, the \textit{n}-gon does have a well-defined boundary, set by the ribbons centerlines. We define the \textit{integrated} Gauss curvature of the unit cell,  $\mathcal{K}_n=\int_{\text{A}} K\,\text{d}A$, where $A$ is a surface enclosed by and tangent to the centerlines of the \textit{n}-gon (for example, the shaded region in Fig.~\ref{fig:2}(b)). The remarkable Gauss-Bonnet theorem~\cite{Stoker:1969vt} states that $\mathcal{K}_n$ is independent of the embedding of the surface, $A$, and it can be determined directly by quantifying the \textit{n}-gon boundary:
\begin{equation}
    \label{eq:gb}
    \mathcal{K}_n = (2-n)\pi - \sum_{i=1}^{n}\kappa_{\text{g}}^i + \sum_{i=1}^{n}\theta^i,
\end{equation}
where $\kappa_{\text{g}}^i=\int_{i} k_\text{g}\,\text{d}s$ is the integrated geodesic curvature of the $i$-th edge and $\theta^i$ is the $i$-th interior angle of the \textit{n}-gon at each crossings (see schematic definitions in Fig.~\ref{fig:2}b). $\mathcal{K}_n$ is the key quantity that we investigate throughout this study. Toward evaluating $\mathcal{K}_n$, as presented in Fig.~\ref{fig:2}(d), we first measured experimental and simulated averages of both the integrated geodesic curvature, $\langle\kappa_\text{g}^\circ \rangle=\frac{1}{n}\sum_{i=1}^{n} \kappa_{\text{g}}^i$, and the interior angles, $\langle \theta^\circ \rangle=\frac{1}{n}\sum_{i=1}^{n} \theta^i$, for representative unit cells with $3 \le n \le 9$.
Within the same cell, we find that all ribbons share the same values of $\kappa_{\text{g}}^i$ and $\theta^i$ (their standard deviation is smaller than the symbol size), as expected from rotational symmetry. In the region $n \ge 5$, $\langle \theta^\circ \rangle \approx 2\pi/3$ independently of \textit{n}, indicating that the exterior triangles remain developable and the Gaussian curvature concentrates at the \textit{n}-gon ~\cite[S2.2]{SupInf_PRL}. To rationalize this observation, we use geometrical arguments based on either a spherical or a conical underlying geometry to estimate the bending energy of the unit cells~\cite[S2.3]{SupInf_PRL}, showing that it is a decreasing function of $n$. Therefore, it is energetically more favorable to bend an \textit{n}-gon with $n>3$ than the external triangles, which can be regarded as nearly developable. However, our current understanding does not draw a full picture of the intricate coupling between elasticity and geometry in these woven structures, whose structural rigidity remains an open problem. Moreover, the vanishing integrated geodesic curvature $\langle \kappa_\text{g}^\circ \rangle \approx 0$ is a direct consequence of the mechanics of elastic ribbons; ribbons favor out-of-plane (instead of in-plane) deformation~\cite{dias2015wunderlich}. Combining these observations with Eq.~\eqref{eq:gb}, we arrive at the expression for the integrated Gauss curvature of a unit cell with straight ribbons:
\begin{equation}
    \label{eq:topologicalcharge}
    \mathcal{K}^\circ_n = \frac{\pi}{3} (6-n),
\end{equation}
where the integer $q^\circ=6-n$ is analogous to the topological charge in curved crystallography~\cite{nelson1983order}. From the measurements on the \textit{n}-gon boundary presented in Fig.~\ref{fig:2}(d), we compute $\mathcal{K}^\circ_n$ as a function of $q^\circ$, as shown in Fig.~\ref{fig:2}(e), onto which we superpose the predictions from Eq.~\eqref{eq:topologicalcharge} and FEM-computations for unit cells with $q^\circ = \{-3,-2,\cdots,3\}$, corresponding to $n=\{9,8,\cdots,3\}$. Despite some discrepancies for $n \le 4$, Eq.~\eqref{eq:topologicalcharge} describes both the experimental and FEM results well. Eq.~\eqref{eq:topologicalcharge} demonstrates that the discrete nature of $q^\circ$ constrains strongly the possible shapes of the unit cells in traditional weaving~\cite{ayres2018beyond, martin2015basketmaker}.

\begin{figure}[t]
    \centering
    \includegraphics[width=\columnwidth]{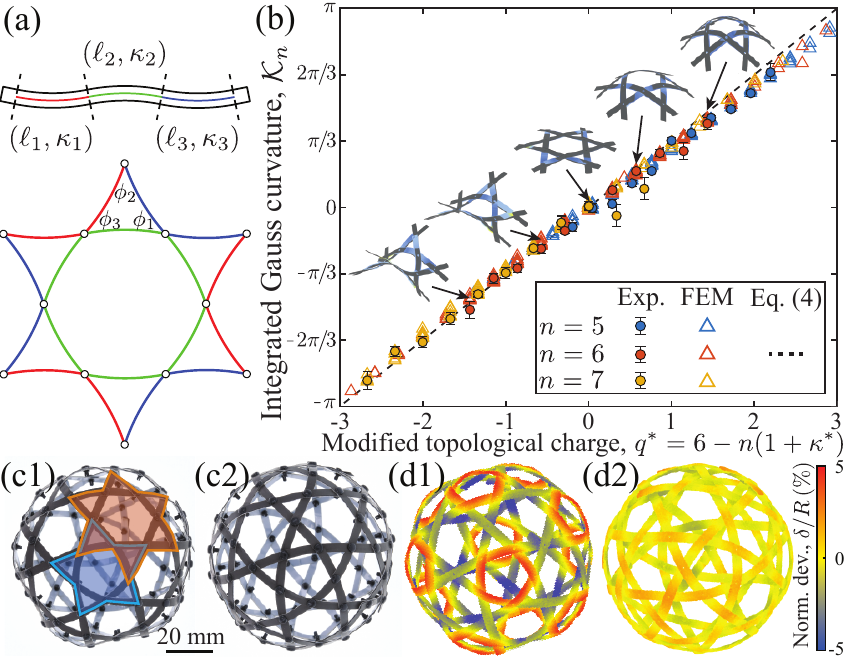}
    \caption{Weaving with curved ribbons. (a)~Schematics of the ribbons (top) and the planar representation of a typical unit cell. The ribbons have 3 distinct curved segments of arc length $\ell_j$ and normalized curvature $\kappa_j$ ($j=1,2,3$; color-coded as red, green, and blue, respectively). (b)~Integrated Gaussian curvature of the unit cells versus $q^*=6-n(1+\kappa^*)$, with $\kappa^*$ from Eq.~\eqref{eq:kappastar}.  {Inset: unit cells with $n=6$, $\kappa_1=\kappa_3=0$, and from left to right $\kappa_2=\{0.5,0.2,0,-0.2, -0.5\}$.} (c1)-(c2)~Photographs of spherical weaves with (c1) straight and (c2) curved ribbons.  {The weaves consist of twelve pentagonal (blue shaded region) and twenty hexagonal (orange shaded region) unit cells.} (d1)-(d2)~Reconstructed $\mu$CT images of the weaves in (c1) and (c2), respectively. The colorbar indicates the normalized voxel-wise radial deviation, $\delta/R$, from a sphere of radius $R=42\,\text{mm}$.}
    \label{fig:3}
\end{figure}

Next, we come back to the non-traditional case of weaving unit cells with naturally curved ribbons, representative examples of which were presented in Fig.~\ref{fig:1}. In Fig.~\ref{fig:3}(a), we show schematic diagrams of an individual curved ribbon (top), as well as the planar representation of the corresponding unit cell (bottom). Note that, although the schematic is drawn planar in Fig.~\ref{fig:3}(a), the woven unit cells are, in general, nonplanar. Each segment with $j=\{1,\,2,\,3\}$ is color-coded as red, green, and blue, respectively. We seek to evaluate the effect of the initial in-plane curvature, $k_j$, on the integrated curvature of the cell, $\mathcal{K}_n$, as a function of \textit{n}. For convenience, we normalize the segment curvature by its arc length; $\kappa_j = k_j \ell_j$. The $(2\pi/n)$-fold rotational symmetry is ensured naturally by the definition of the unit cell when \textit{n} is even, and enforced when \textit{n} is odd by further imposing $\ell_1=\ell_3$ and $\kappa_1=\kappa_3$. This dense sampling is sufficient to quantify the effect of the in-plane geometry of the ribbons. Motivated by our findings for unit cells with straight ribbons (Fig.~\ref{fig:2}d), we make the following remarks. First, we assume that the ribbons keep their in-plane curvature when woven; hence, $\kappa_\text{g}^j=\int_j k_\text{g}\,\text{d}s = \int_j k_j\,\text{d}s = k_j \ell_j = \kappa_j$ for every segment of the \textit{n}-gon~\cite{dias2015wunderlich}. Second, for straight ribbons, we found that the outer triangles remained isometric, thereby enclosing a surface of vanishing integrated curvature; a statement that we now assume to remain valid for unit cells woven with curved ribbons. {Leveraging this assumption on the outer triangles}, we evaluate the interior angles of the triangles (opposite to the arc $\ell_j$; see Fig.~\ref{fig:3}a) using Euclidean trigonometry: $\phi_j = \cos^{-1} \bigg( \frac{\sum_{m\neq j} (g_m^2  - g_j^2) }{2 \prod_{m\neq j} g_k}  \bigg) -\sum_{m\neq j}\frac{\kappa_m}{2}$, where $g_j = 2 \sin (\kappa_j/2)\cdot \ell_j/\kappa_j $. In turn, the interior angles of the \textit{n}-gon, $\theta^i$, are the supplementary angle of either $\phi_1$ or $\phi_3$; such that $\sum_{i=1}^n \theta^i= \{n(\pi-\phi_1) + n(\pi-\phi_3)\} / 2$. Thus, using the Gauss-Bonnet theorem stated in Eq.~\eqref{eq:gb}, the integrated curvature of a unit cell reads $\mathcal{K}_n = \frac{\pi}{3} [6 -n(f+\kappa^*)]$, where $f = \frac{3}{2\pi}\cos^{-1} \big( \frac{ g_2^2 - g_1^2 - g_3^2 }{2 g_1 g_3} \big)$, and
\begin{equation}
    \label{eq:kappastar}
    \kappa^* = \frac{3}{4\pi}(-\kappa_1+2\kappa_2-\kappa_3).
\end{equation}
The arc length $\ell_j$ and the curvature $\kappa_j$ are coupled through the nonlinear term $f$. Noting that $g_j \approx \ell_j$ in the range of in-plane curvatures considered, $|\kappa_j| \le 0.5$ (\textit{e.g.}, $g_j (\kappa_j=0.5) \approx 0.99\,l_j$), we take the asymptotic limit of $|\kappa_j| \ll 1$. We further impose $\ell_1=\ell_2=\ell_3$ to quantify only the effect of the in-plane curvatures. Ultimately, we obtain $f = 1$ and Gauss-Bonnet reduces to
\begin{equation}
    \label{eq:K_n_curved_linearized}
    \mathcal{K}_n(\kappa_1,\,\kappa_2,\,\kappa_3) = \frac{\pi}{3}[6 - n(1+\kappa^*)].
\end{equation}
From the similitude between Eqs.~\eqref{eq:K_n_curved_linearized} and ~\eqref{eq:topologicalcharge}, we define $q^*=6-n(1+\kappa^*)$ as the \textit{modified topological charge} of the unit cell with \textit{curved} ribbons. We highlight that $q^*$ can be varied smoothly using curved ribbons, yielding a continuous range of $\mathcal{K}_n$,
whereas $\mathcal{K}^\circ_n$ in Eq.~\eqref{eq:topologicalcharge} was discrete and restricted to multiples of $\pi/3$.

\begin{figure*}[t]
    \centering
    \includegraphics[width=\textwidth]{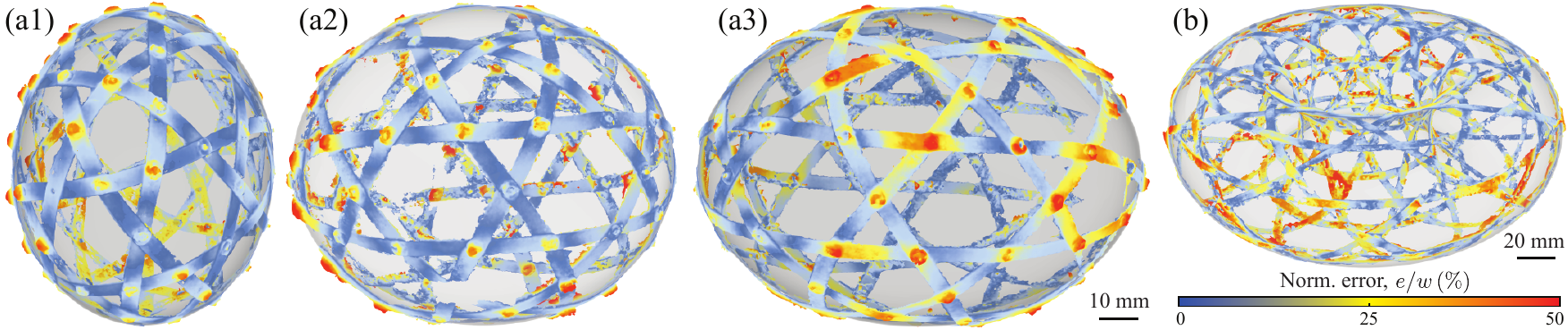}
    \caption{Nonspherical weaves with initially curved ribbons, reconstructed by photogrammetry.  {The colorbar indicates the normalized deviation from a target surface, $e/w$, where $w$ is the ribbon width.} (a1)-(a3)~Ellipsoidal weaves of aspect ratios, $a/b=\{0.75,1.25,1.5\}$, respectively. (b)~Toroidal weave of inner radius $r_\text{i}=35\,\text{mm}$ and outer radius $r_\text{o}=105\,\text{mm}$. The planar geometries of the underlying curved ribbons for each of these weaves are provided in~\cite[S3.2-3]{SupInf_PRL}} 
    \label{fig:4}
\end{figure*}

In Fig.~\ref{fig:3}(b), we plot experimental and FEM data for $\mathcal{K}_n$ \textit{vs.} $q^*$, while fixing $\ell_j=15\,\text{mm}$, to systematically explore the parameter space $n=\{5,6,7\}$ and $\kappa_2=\{-0.5,\cdots,0.5\}$ (in steps of 0.1). Also, the experiments had $\kappa_1=\kappa_3=0$ (total of 33 configurations, with two experiments per configuration) and the simulations had $\kappa_1=\pm \kappa_3=\{-0.5,-0.2,0,0.2,0.5\}$ (total of 352 configurations). To convey the change of shape associated to variations of the integrated curvature, we also juxtapose the unit cells with $n=6$ and varying ribbon curvatures that we presented in Fig.~\ref{fig:1}. As above for the unit cells with straight ribbons, we measured $\theta^i$ and $\kappa_{\text{g}}^i$ of the \textit{n}-gon and used Eq.~\eqref{eq:gb} to compute $\mathcal{K}_n$~\cite[S1.3]{SupInf_PRL}. Remarkably, we find that the data in Fig.~\ref{fig:3} collapse over the full range of $-\pi \le \mathcal{K}_n \le \pi$. This continuous variation for curved ribbons contrasts to the analogous result for traditional weaving (Fig.~\ref{fig:2}e), where $\mathcal{K}^\circ_n$ was limited to discrete steps of $\pi/3$ (\textit{cf.} Fig.~\ref{fig:2}e). Importantly, our geometric prediction for $\mathcal{K}_n$ from Eq.~\eqref{eq:K_n_curved_linearized} is in excellent agreement with the data, demonstrating that in-plane ribbons geometry is at the heart of our triaxial weaving problem.

Thus far, we followed a unit-cell approach to demonstrate that smooth weaving can be physically realized with curved ribbons, purely from geometric design principles. We now seek to assemble these unit cells into a spherical weave, adopting the topology of the \textit{rectified truncated icosahedron}~\cite{atI1, atI2} for the layout of our design. We fix the segment length to $\ell_\text{s}=15\,\text{mm}$ and inject a (nondimensional) segment curvature $\kappa_\text{s}=k_\text{s} \ell_\text{s}$ into the pentagonal cells. The resulting weave comprises 12 pentagonal cells with  {$(\kappa_1,\kappa_2,\kappa_3)=(0,\kappa_\text{s},0)$ and 20 hexagonal cells with $(\kappa_1,\kappa_2,\kappa_3)=(\kappa_\text{s},0,0)$, marked with blue and orange shaded regions in Fig.~\ref{fig:3}(c1), respectively.} In Figs.~\ref{fig:3}(c1)-(c2), we present photographs of two spherical weaves: one with straight ribbon ($\kappa_\text{s}=0$), the traditional case, and the other with curved ribbons ($\kappa_\text{s}=0.3$). The corresponding $\mu$CT images are shown in Figs.~\ref{fig:3}(d1)-(d2), color-coded by the radial distance between the scans and a sphere of radius $R=42\,\text{mm}$, $\delta\,[\text{mm}]$. Negative values of $\delta$ indicate voxels located inside the targeted sphere. For the weave with straight ribbons (Fig.~\ref{fig:3}(d1)), the pentagons protrude from the reference sphere, with 5\,\% maximum radial deviation. This faceted geometry is a signature of the localized curvature intrinsic to the discrete nature of traditional weaving;  Eq.~(\ref{eq:K_n_curved_linearized}), predicts $\mathcal{K}_5=\pi/3$ for the pentagons and $\mathcal{K}_6=0$ for the hexagons. By contrast, the weave with curved ribbons ($\kappa_\text{s}=0.3$) shown in Fig.~\ref{fig:3}(d2) exhibits a significantly smoother shape, with a radial deviation within 1\,\% of the perfect sphere; Eq.~(\ref{eq:K_n_curved_linearized}) predicts $\mathcal{K}_5 = 0.3$ and $\mathcal{K}_6 = 0.45$.

Our unit cells with curved ribbons are rotationally symmetric. Hence, the possible design space available by their tessellation is limited to shapes with local symmetry (\textit{e.g.}, the sphere in Fig.~\ref{fig:3}c). We do not expect this approach to be, in general, viable to design weaves with more complex or arbitrary geometries. To overcome this limitation, we expanded our framework to design the initial shape of piecewise-circular ribbons that are to be woven into a given target surface. Similar to what we did for the unit cells, the injection of geodesic curvature into the weave through the in-plane curvature of the ribbons is at the core of the procedure. Our design protocol (detailed in \cite[S3.1]{SupInf_PRL}) consists of inputting a \textit{target surface}, onto which we project a \textit{graph} representing the triaxial weave topology. This graph contains \textit{nodes} (corresponding to the crossing points of the ribbons) and \textit{edges} for their connectivity. At each node, a geodesic turning angle between consecutive nodes is computed with respect to the target surface. The shape of piecewise-circular segments of the ribbon is then obtained by averaging the two geodesic turning angles from its neighboring crossing points.

In Figs.~\ref{fig:4}(a1-a3), as a first example of nonspherical designs, we present reconstructed images of ellipsoidal weaves of an equatorial radius, $b=40\,\text{mm}$, and polar radii, $a=\{30,50,60\}\,[\text{mm}]$. The graph for these weaves was obtained by adopting the topology of the \textit{rectified truncated icosahedron}~\cite{atI1, atI2} and linearly expanding it by a factor $a$ along the $x,y$-axes, and a factor $b$ along the $z$-axis~\cite[S3.2]{SupInf_PRL}. Photogrammetry was used to reconstruct the 3D shape of the ellipsoidal woven structures from a series of photos with different perspectives~\cite[S1.4]{SupInf_PRL}. The reconstructed meshes are color-coded with errors from their target surface normalized by their ribbon width, $w = 4\,\text{mm}$. Even though our design strategy did not incorporate the elastic energy of the ribbons, excellent agreement is found between the target surfaces and the weaves, exhibiting the distance error smaller than 50\,\% of the ribbon width. In an ellipsoid, only the meridians and the equator are the closed geodesics~\cite{Stoker:1969vt}, so it is not trivial not achieve smooth oblate (Fig.~\ref{fig:4}.a1) and prolate (Figs.~\ref{fig:4}.a2-a3) ellipsoidal weaves. Curved ribbons enable to accommodate variations of the aspect ratio of the ellipsoids while keeping the same weave topology. As a second example, a smooth torus (genus-zero surface with zero total curvature~\cite{Stoker:1969vt}) cannot be achieved through traditional weaving; using straight ribbons inevitably requires the placement of pentagonal and hexagonal defects, albeit with a localization of curvature that leads to faceted geometry. By contrast, as demonstrated by the physical realization in Fig.~\ref{fig:4}(b), our design with curved ribbons yields a toroidal weave with hexagonal cells alone, the in-plane curvature of the ribbons distributing the total curvature. The presented smooth toroidal weave has an inner radius $r_\text{i}=35\,\text{mm}$ and an outer radius $r_\text{o}=105\,\text{mm}$. The topology of the toroidal weave was obtained by mapping a regular triaxial pattern in a 2D parameter space to a 3D toroidal target surface~\cite[S3.3]{SupInf_PRL}. 

The planar layout of the curved ribbons used in the above designs (spherical, ellipsoidal, and toroidal) are provided in~\cite[]{SupInf_PRL} and can be cut and woven by the interested reader.

Our work demonstrates that the discrete nature of traditional triaxial weaving can be circumvented by using initially curved, piecewise-circular ribbons. The shape of the weaves can be decoupled from their topology, with multiple topological layouts and ribbons geometries leading to the same woven shapes. However, when woven, these geometrically identical solutions store elastic energy differently~\cite[S2.1]{SupInf_PRL}. This observation calls for a full optimization problem, which we hope future work will address, where both the distance from the target surface and the associated elastic energy are minimized in tandem by changing the geodesic curvatures and the segment lengths of the ribbon as design parameters. 
Beyond art and architecture, future implementations of our design framework may include morphing structures in which the in-plane curvature of ribbons would be pre-programmed into the ribbons and actuated upon stimuli to attain desired target shapes. 

\begin{acknowledgments}
\noindent We thank Yingying Ren, Julian Panetta, Florin Isvoranu, Christopher Brandt, and Mark Pauly for fruitful discussions.
\end{acknowledgments}

\end{document}


\centerline{\bf \large Smooth triaxial weaving with naturally curved ribbons} 
\vspace{5mm}
\centerline{\bf \large -- Supplemental Information -- }
\vspace{5mm}
\centerline{\large Changyeob Baek, Alison G. Martin, Samuel Poincloux, Tian Chen, and Pedro M. Reis}

\author{Changyeob Baek, Alison G. Martin, Samuel Poincloux, Tian Chen, and Pedro M. Reis}
\vspace{0mm}

\section{Image processing of the unit cells}
\label{sec:imageprocessing}

In this section, we detail the image-processing algorithm that we developed to process the volumetric images of the unit cells obtained through X-ray micro-computed tomography ($\mu$CT). The goal is to construct a centerline-based description of each of the ribbons within the. Our image-processing procedure involves the following four steps: (i) acquisition of the $\mu$CT volumetric image of the unit cell, (ii) segmentation of the voxels corresponding to each ribbon, (iii) extraction of the centerline, (iv) refinement and smoothing of the centerline, and (v) quantification of the material frame of the ribbon. As a result, we obtain a centerline-based description of the ribbons composing the unit cell. This centerline data is then used to calculate the geodesic curvature of the ribbons and interior angles of the unit cell.

The framed-centerline of the $i$-th ribbon is denoted as $\Gamma^i = \{\bm{\gamma}^i; \mathbf{t}^i,\,\mathbf{n}^i,\,\mathbf{b}^i\}$, where the superscript $i$ corresponds the index of each ribbon in the unit cell. Specifically, $\bm{\gamma}^i$ is a three-dimensional (3D) discrete curve containing $N^i$ vertices; $\bm{\gamma}^i = \{\bm{\gamma}^i_j\}$ $(j=1,\,2,\,\cdots,\,N^i)$ The subscript $j$ denotes the index of the vertices. Each point along the centerline $\bm{\gamma}^i_j$ is assigned with the material frame $(\mathbf{t}^i_j,\,\mathbf{n}^i_j,\,\mathbf{b}^i_j)$ where $\mathbf{t}^i_j$ is the tangent vector, $\mathbf{n}^i_j$ is the normal vector, and $\mathbf{b}^i_j$ is the binormal vector. Ultimately, based on the complete description of the centerlines of the ribbons, we compute the following two quantities: (a) the $i$-th interior angle of the $n$-gon, $\theta^i$, and (b) the integrated geodesic curvature of the $i$-th segment of the $n$-gon, $\kappa^i = \int_{i} \kappa_\mathrm{g}\,\mathrm{d}s$, of the unit cell. With the quantification of $\theta^i$ and $\kappa^i$ at hand, we can then readily compute the integrated Gauss curvature of the unit cells thanks to the Gauss-Bonnet theorem presented in Eq.~(1) of the main text.

Next, in Sec.~\ref{subsec:imageprocessing_1}, we will first describe the experimental fabrication of the ribbons and image acquisition protocol using the $\mu$CT. The fabrication protocol was developed specifically so as to facilitate the $\mu$CT imaging and the extraction of the centerlines of the ribbons. In Sec.~\ref{subsec:imageprocessing_2}, we then detail an algorithm that we developed in-house to quantify $\Gamma^i$. Finally, in Sec.~\ref{subsec:imageprocessing_3}, we compute $\theta^i$ and $\kappa^i$ based on $\Gamma^i$.

\subsection{Acquisition of the volumetric $\mu$CT images}
\label{subsec:imageprocessing_1}

\begin{figure}[h!]
    \centering
    \includegraphics[width=0.8\columnwidth]{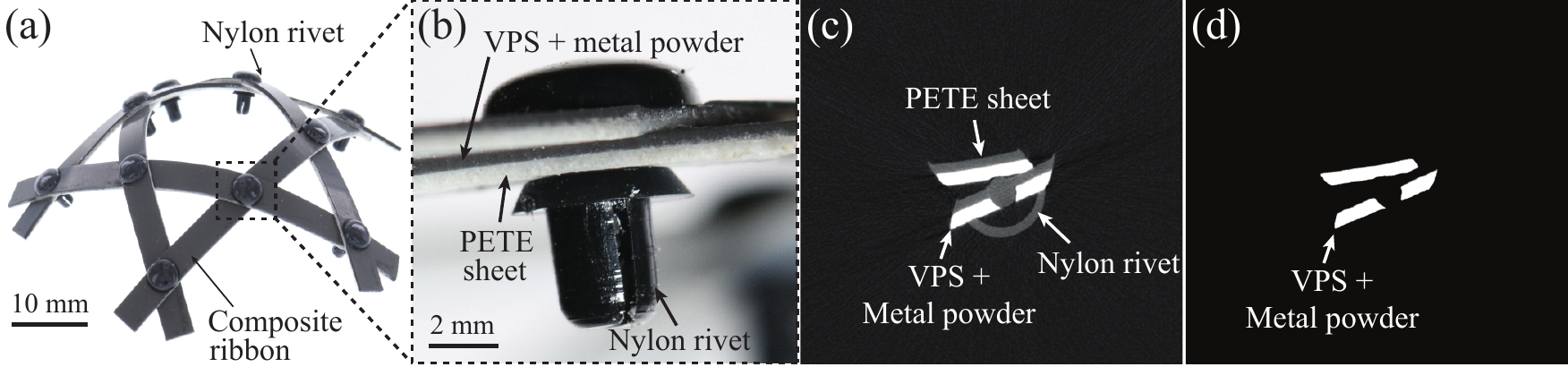}
    \caption{(a)~Photograph of a representative unit cell woven with composite ribbons, which consist of a bilayer of polyethylene terephthalate (PETE) and metal-infused vinylpolysiloxane (VPS). The ribbons were assembled into the unit cell using nylon rivets. (b)~Magnified view of a crossing region between two ribbons. (c)~Cross-sectional view of the crossing region obtained from a $\mu$CT scan. In the X-ray images, the metal-infused VPS layer appears brighter than any other object in the image due to its high radiopacity. (d)~The $\mu$CT image can be brightness-adjusted to only shows the metal-infused VPS layer on an otherwise dark background.}
    \label{fig:weave_fabrication}
\end{figure}

In the main text, we introduced the fabrication procedure of the unit cells, which comprise $n$ composite ribbons. These bilayer ribbons consist of a layer of metal-infused vinylpolysiloxane (VPS-16, Zhermack) and a layer of polyethylene (PETE, Plastic Shim Pack DM1210, Partwell Group). We assembled the composite ribbons using nylon rivets (Snap Rivet \SI{4.2}{\milli\metre}, Distrelec AG, Switzerland). In Fig.~\ref{fig:weave_fabrication}(a) and (b), we present a photograph of the unit cell and a magnified view of the crossings. In Fig.~\ref{fig:weave_fabrication}(c), we present a cross-sectional view of the $\mu$CT image, zooming in the vicinity of the crossing of two ribbons. As stated in the main text, the higher radiopacity of the metal-infused VPS layers enables us to segment the VPS layer. In Fig.~\ref{fig:weave_fabrication}(d), we present an image whose brightness has been adjusted manually to highlight only the VPS regions. 

\subsection{Quantifying the centerlines of the ribbons}
\label{subsec:imageprocessing_2}

\begin{figure}[b]
    \centering
    \includegraphics[width=0.8\columnwidth]{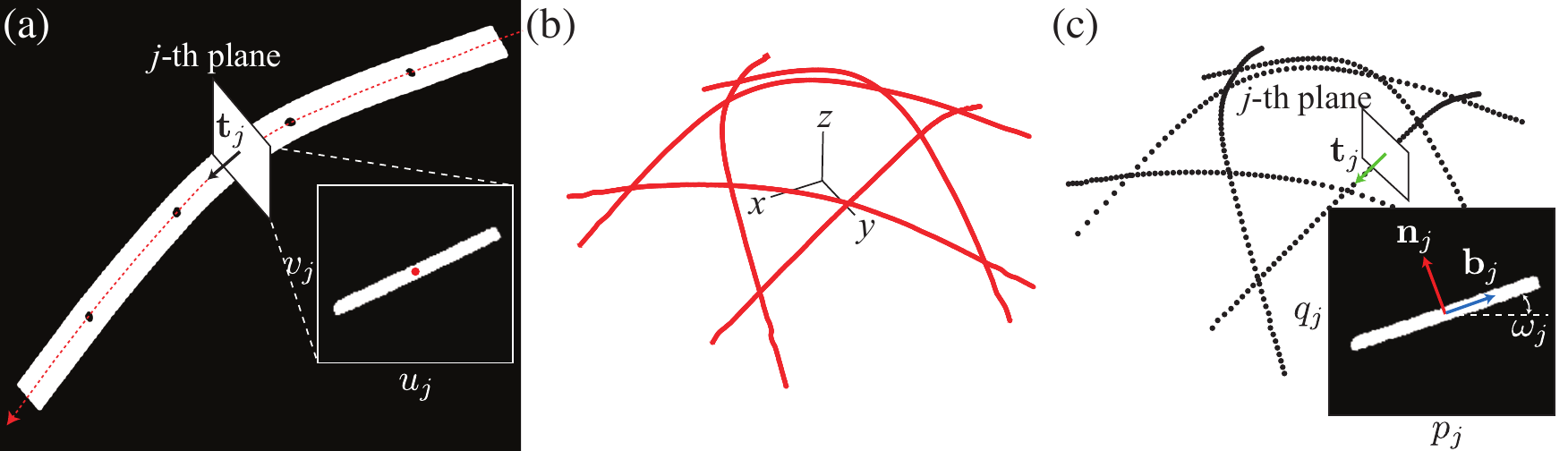}
    \caption{(a)~A reconstructed $\mu$CT image of an isolated ribbon. We investigate a cross-sectional cut at $\bm{\gamma}_j$ and quantify its centroid. (b)~Plot of the refined centerlines of the full unit cell. (c)~Having obtained a smoothed centerline (black dots), we march through $\bm{\gamma}$, along its tangent ($\mathbf{t}$; green arrows), and investigate its cross-sectional cut (inset) to obtain the material frame.}
    \label{fig:weave_imageprocessing}
\end{figure}

Since the PETE layers physically separate the metal-infused VPS layers (see Fig.~\ref{fig:weave_fabrication}d), we were able to isolate volumetric images of every single ribbon from the rest of the image using the command \texttt{bwconncomp} in MATLAB. Fig.~\ref{fig:weave_imageprocessing}(a) shows a reconstructed image of a single ribbon, from which we extract the framed centerline, $\Gamma^i$, using the procedure described in this section.

We start by manually inputting an ansatz of the centerline, $\bm{\gamma} = \{\bm{\gamma}_j\}$, depicted by the red dashed line in Fig.~\ref{fig:weave_imageprocessing}(a). We will temporarily omit the superscript $i$ for the ribbon index, for the sake of brevity. To set this ansatz, we first picked a few seeding points (the number of manual input is approximately ten) on the volumetric image using the MATLAB command \texttt{datacursormode} and interpolated those points using the MATLAB command \texttt{fit}. Note that $\bm{\gamma}=\{\bm{\gamma}_j\}$ is not yet the final centerline; the components of $\bm{\gamma}$ are still to be iterated for a more accurate representation of the actual centerline. The ansatz $\bm{\gamma}$ is a 3D polygonal curve comprising $N$ vertices; $\bm{\gamma}=\{\bm{\gamma}_j\}~(j=1,\,2,\,\cdots,\,N)$. The initial value of $N$ is determined from a combination of the resolution of the $\mu$CT tomographic imaging ($a$) and the total length of the ribbon ($\ell$). Throughout our study, we used $a=\SI{29.3}{\micro\metre}$ and $\ell \simeq \SI{30}{\milli\metre}$ (thus $N \simeq 1,000$) to set the initial guess of the centerline $\bm{\gamma}$ along the ribbon.

Having identified the ansatz of the centerline ($\bm{\gamma}$), we updated $\bm{\gamma}$ with a refined centerline that is obtained as follows. At each $j$-th point on the ansatz $\bm{\gamma}_j$, we extracted a slice-view of the volumetric image along a plane normal to the tangent of the ansatz at the $j$-th point, $\mathbf{t}_j \equiv \bm{\gamma}_{j+1} - \bm{\gamma}_{j}~(j=1,\,2,\,\cdots,\,N-1)$ (see inset of Fig.~\ref{fig:weave_imageprocessing}a). By definition, this cutting plane lies on the null-space of $\mathbf{t}_{j}$. For convenience, we denote the basis of this 2D null-space as $\{\mathbf{u}_j,\,\mathbf{v}_j\}$. We then obtained the centroid of the slice-view image in $(u_j,\,v_j)$, using the MATLAB command \texttt{regionprops} (red point in the inset of Fig.~\ref{fig:weave_imageprocessing}a), and updated the centerline as
\begin{equation}
\label{eq:ansatz_to_refined}
     \bm{\gamma}_j + u_j \mathbf{u}_j + v_j \mathbf{v}_j \rightarrow  \bm{\gamma}_j.
\end{equation}
The resulting centerline of the ribbon is depicted by the red solid lines in Fig.~\ref{fig:weave_imageprocessing}(b). During this stage of refinement of the centerline, the number of discretization points, $N$, remains unchanged from that of the centerline ansatz.

The refined centerline $\bm{\gamma} = \{\bm{\gamma}_j\}$ would serve as the `\textit{exact}' centerline of the ribbon in the limit of infinitely high resolution of the X-ray scan. However, due to the finite, even if excellent, resolution of our $\mu$CT, the refined centerline is still too noisy to be used for computing derivative quantities (\textit{e.g.}, curvatures). Specifically, the number of voxels along the thickness of the ribbon is approximately $t_\text{c}/a= 350\,\mu \text{m}/29.3\,\mu \text{m} \simeq 10$ for the unit cells that we investigated, which is not sufficiently high to prevent numerical errors from the voxelization process. Therefore, additional smoothing of the centerline is necessary. As such, we perform a window-average $\bm{\gamma}$ in every $m$ steps:
\begin{equation}
\label{eq:refined_to_final}
    \frac{1}{m} \sum_{k=m(j-1)+1}^{mj} \bm{\gamma}_k \rightarrow \bm{\gamma}_j. 
\end{equation}
During this averaging step, the level of discretization of the centerline was decreased from $N$ to $N/m$. In Fig.~\ref{fig:weave_imageprocessing}(c), we present an example of the final centerline, depicted by the black dotted lines. For the analysis of the unit cells, we set $m=15$; thus, a total of $N \simeq 70$ data points represent each centerline.

Having obtained the final centerline $\bm{\gamma} = \{\bm{\gamma}_j\}$, we computed the corresponding material frame at each point on the centerline. The first components of the frame, $\mathbf{t}_j$ (tangent of the centerline), is computed from the numerical tangent of $\bm{\gamma}$ via the relation $\mathbf{t}_j=(\bm{\gamma}_{j+1} - \bm{\gamma}_{j})/|\bm{\gamma}_{j+1} - \bm{\gamma}_{j}|$. In order to obtain the other two components of the material frame $(\mathbf{n}_j,\,\mathbf{b}_j)$, we investigated a cross-sectional view of the image along a plane of normal $\mathbf{t}_j$ with a null-space of the basis $(\mathbf{p}_j,\,\mathbf{q}_j)$ (see inset of Fig.~\ref{fig:weave_imageprocessing}c). We then obtained the angle between the $p$-axis and the principal orientation of the $j$-th cross-sectional view, $\omega_j$, using the MATLAB command \texttt{regionprops}. The material frame $(\mathbf{n}_j,\,\mathbf{b}_j)$ is then determined by rotating the basis $(\mathbf{p}_j,\,\mathbf{q}_j)$ with $\omega_j$. Finally, the material frame at the $j$-th point of the centerline, $\bm{\gamma}_j$, is expressed as
\begin{equation}
    \begin{aligned}
    \mathbf{t}_j &= \frac{\bm{\gamma}_{j+1} - \bm{\gamma}_{j}}{|\bm{\gamma}_{j+1} - \bm{\gamma}_{j}|}, \\
    \mathbf{n}_j &= -(\sin \omega_j )\mathbf{p}_j + (\cos \omega_j )\mathbf{q}_j, \\
    \mathbf{b}_j &= (\cos \omega_j )\mathbf{p}_j + (\sin \omega_j )\mathbf{q}_j.
    \end{aligned}
\end{equation}
A representative example of the final centerline, together with the corresponding material frame, was presented in Fig.~2(b) of the main text.

\subsection{Measurement of $\theta^i$ and $\kappa_{\text{g}}^i$}
\label{subsec:imageprocessing_3}

Having obtained the framed centerlines $\Gamma^i = (\bm{\gamma}^i;\, \mathbf{t}^i,\,\mathbf{n}^i,\,\mathbf{b}^i)$ from the $\mu$CT images, we can now quantify both the $i$-th interior angle of the $n$-gon, $\theta^i$, and the integrated geodesic curvature of the $i$-th edge of the $n$-gon, $\kappa_{\text{g}}^i=\int_{i} k_\text{g}\,\text{d}s$. These quantities will be used directly for the calculation of the integrated Gauss curvature of the unit cells through the Gauss-Bonnet theorem in Eq.~(1) of the main text. First, $\theta^i$ is obtained by measuring the angle between a pair of centerlines involved in the $i$-th crossing. Second, $\kappa_{\text{g}}^i$ is measured by summing up the discrete geodesic curvature of the $i$-th centerline along the $i$-th edge of the $n$-gon. The integrated geodesic curvature along the framed centerline $\{\bm{\gamma}^i_j\}$ between the points $j=a$ and $j=b$ ($1 < a < b < N^i$) can be expressed as~\cite{bobenko2008discrete, bergou2008DER}:
\begin{equation}
    \label{eq:weave_geodturn}
    \kappa_{\text{g}}^i = \sum_{j=a}^{b} \bigg[ \bigg( \frac{2 \mathbf{t}^i_{j-1} \times \mathbf{t}^i_{j}}{ |\mathbf{t}^i_{j-1}| |\mathbf{t}^i_{j}| + \mathbf{t}^i_{j-1} \cdot \mathbf{t}^i_{j} } \bigg) \cdot \mathbf{n}^i_j \bigg].
\end{equation}
Specifically, the term inside the parenthesis in Eq.~\eqref{eq:weave_geodturn} quantifies the discrete curvature vector of the $j$-th vertex. The component of the discrete curvature along the normal vector $\mathbf{n}_j$ is the discrete geodesic curvature. Since it is already an integrated quantity (hence, dimensionless), we obtain the integrated geodesic curvature by summing the term inside the square brackets in Eq.~\eqref{eq:weave_geodturn}, from $j=a$ to $j=b$.

\begin{figure}[h!]
    \centering
    \includegraphics[width=0.9\columnwidth]{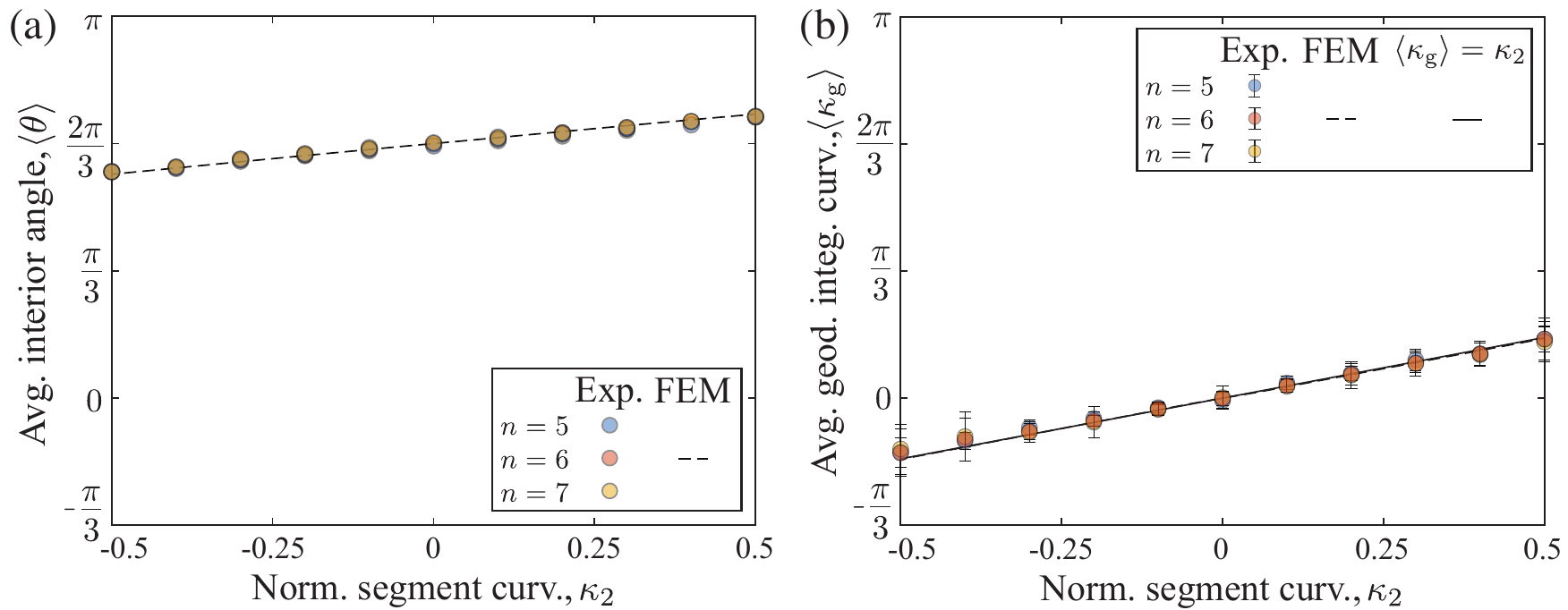}
    \caption{(a)~Average interior angle, $\langle\theta\rangle=\frac{1}{n}\sum_{i=1}^n \theta^i$, and (b)~Average integrated geodesic curvature, $\langle\kappa_\text{g}\rangle=\frac{1}{n}\sum_{i=1}^n \kappa_{\text{g}}^i$, as functions of the nondimensional segment curvature, $\kappa_2$, for woven unit cells with $n=\{5,\,6,\,7\}$. For the FEM simulations, only the results with $n=6$ are presented. The error bars for $\langle \kappa_\text{g} \rangle$ represent the standard deviation of the $\kappa_{\text{g}}^i$ measurements within a single unit cell.}
    \label{fig:curved_anglegeod}
\end{figure}

In Fig.~2 of the main text, we reported the average interior angle, $\langle \theta \rangle=\frac{1}{n}\sum_{i=1}^n \theta^i$, and the average integrated geodesic curvature, $\langle \kappa_\text{g} \rangle=\frac{1}{n}\sum_{i=1}^n \kappa_{\text{g}}^i$, of representative units cells constructed with initially straight ribbons. In Fig.~\ref{fig:curved_anglegeod}(a), we plot $\langle \theta \rangle$ for unit cells constructed with curved ribbons as a function of the nondimensional segment curvature, $\kappa_2$. We focus on three specific cases of weaves with $n=\{5,\,6,\,7\}$ ribbons. In that same figure, we also plot the data obtained from the corresponding series of FEM simulations (dashed lines), which are in excellent agreement with the experiments. More detail on the FEM simulations is provided below, in Sec~\ref{sec:FEM}. The average interior angles were obtained by averaging the angles in a single unit cell, $\{\theta^1,\,\theta^2,\,\cdots,\,\theta^n\}$, and the associated standard deviation was found to be negligible (smaller than the symbol size in the plot). We recall that, for traditional weaving (straight ribbons), the only possible value of the interior angle was $\langle \theta \rangle=2 \pi/3$, regardless of the value of $n$ (see Fig.~2 of the main text). By contrast, the unit cells with naturally curved ribbons exhibit a continuous range of interior angles by tuning the in-plane geometry of the ribbons.

In Fig.~\ref{fig:curved_anglegeod}(b), we plot the average integrated geodesic curvature along the $n$-gon, $\bar{\kappa}_\text{g}$, as a function of the injected in-plane curvature along the middle segment ($\kappa_2$) of each ribbon. As stated in the main text, $\kappa_2$ contributes directly to the geodesic curvature of the segments of $n$-gon: the amount of injected curvature $\kappa_2 = k_2 \ell_2$ is equal to the measured integrated geodesic curvature $\kappa_{\text{g}}^i=\int_{i} k_g\,\text{d} s$. Again, this finding confirms that the geodesic curvature of the ribbon ($k_\text{g}$) remains unchanged from that of the reference configuration ($k_2$) during deformation, thus yielding $\int_{i} k_g^i \,\text{d} s = \int_{i} k_2\,\text{d} s = k_2 \ell_2$~(black solid line in Fig.~\ref{fig:curved_anglegeod}b). The FEM measurements of $\bar{\kappa}_\text{g}$ do not have any observable standard deviation since the rotational symmetry of the problem is well imposed in the FEM mesh. However, the experimental measurements do have some level of uncertainty, presumably due to imperfections originating from the fabrication process and partly due to the finite resolution of the $\mu$CT scan. The error bars in Fig.~\ref{fig:curved_anglegeod}(b) represent the standard deviation of $\kappa_\text{g}^i~(i=1,\,2,\,\cdots,\,n)$, measured from two independent scans per configuration. In any case, the uncertainty of the experimental measurements remains relatively small (of the order of $\pi/10$), and the experiments and FEM simulations are found to be in excellent agreement.

\subsection{Reconstruction of 3D geometries of larger woven structures}

The ellipsoidal and toroidal weaves presented in Fig. 4 of the main text were too large to fit in the $\mu$CT scanner, which was employed earlier in the study to scan the 3D geometry of the various unit cells. Instead, for these larger structures, we used photogrammetry to reconstruct their 3D shape. Photogrammetry is a technique used to acquire the precise 3D geometric information of an object by assembling multiple two-dimensional images of the sample. In order to systematically obtain a series of photos with prescribed viewing angles, we placed the woven structures on the rotating stage of a professional photo booth (Packshot Creator R3L T 360). A series of photos was taken at regular intervals of longitudinal rotation (every $6^\circ$) around the objects, and two different values of latitude ($45^\circ$ and $80^\circ$). The photos were then post-processed using commercially available software (Agisoft Metashape) to construct a 3D mesh and, consequently, the cloud-point coordinates of the surface of the structure. With these surface coordinates at hand, we could then quantify the deviation of the scanned weaves from their target surface. Even if the spatial resolution of this photogrammetry technique is lower than that of the X-ray tomography, we still found it appropriate to quantify the pointwise radial deviation of the woven structures from their target surfaces.

\section{Mechanical modeling of the unit cells.}

\subsection{FEM simulations of the representative unit cells}
\label{sec:FEM}

\begin{figure}[b!]
    \centering
    \includegraphics[width=0.8\textwidth]{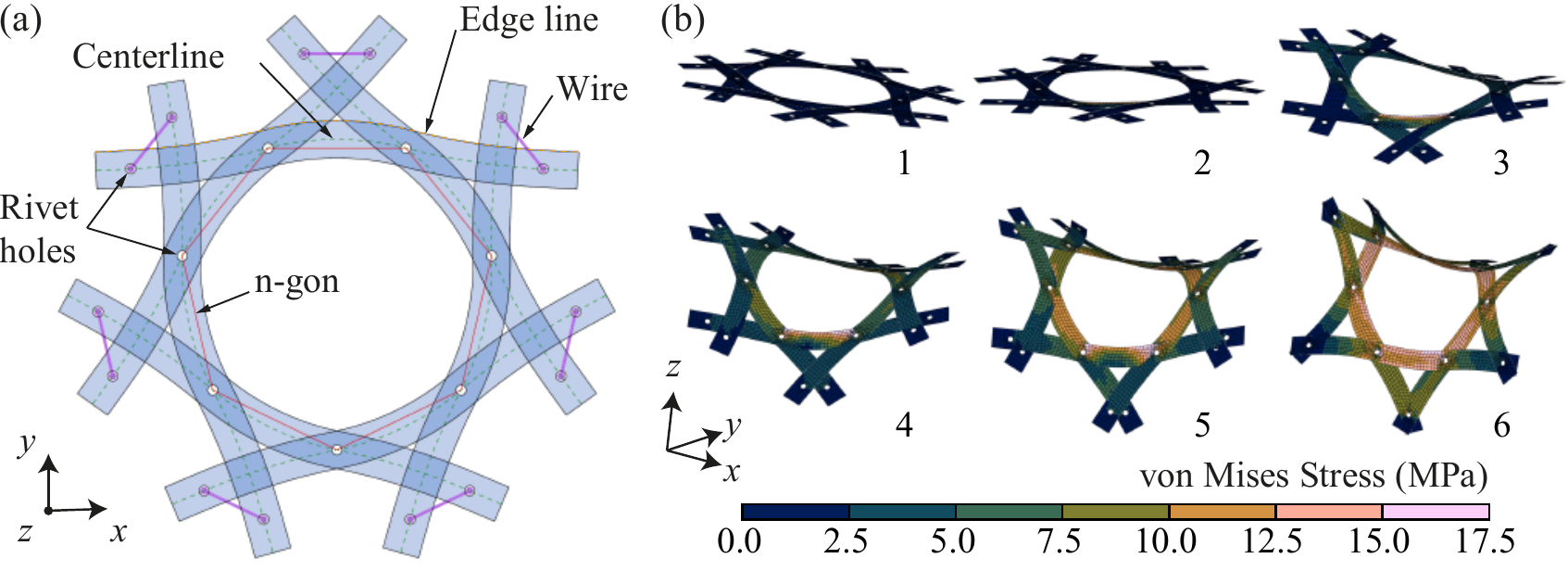}
    \caption{(a) Numerical setup and (b) representative results of the FEM simulations of a unit cell with $n=7$ ribbons, with $\ell=\SI{15}{\milli\metre}$ and $\kappa_j=[0.2,0.5,0.2]$. (a)~Schematic of the model setup. The ribbons are positioned such that they form an inner polygon of $n$ edges, and strings are used to attach the end points. (b)~As the length of the strings is reduced to 0, the initially flat shape becomes curved and exhibits a 3D shape. Von Mises stress were plotted over the deformed meshes.}
    \label{fig:m-fem}
\end{figure}

We followed the Finite Element Method (FEM) to perform simulations of the woven unit cells using the commercial package Abaqus 6.14. Quadratic shell elements were used to model the ribbons with a linear elastic material model ($E=\SI{3000}{\mega\pascal}, \nu = 0.3$). The holes for the rivets were included in the geometry of the ribbons. The initial positions of the ribbons were laid out such that the center segment of the ribbons formed a regular polygon of $n$ edges (see Fig.~\ref{fig:m-fem}a). The vertices of this polygon are the overlapping inner rivet holes. Note that the edge length of this polygon decreases as the absolute value of the curvature of the center segment increases. Each rivet holes (both inner and outer) was kinematically tied in all degrees of freedom (DOFs) to a reference point at the center of the hole. For the overlapping inner rivet holes, the two respective reference points were then kinematically tied in all DOFs such that they preserve planarity.

The $n$ pairs of outer rivet holes do not, in general, overlap at the start of the simulation. Each pair of two corresponding rivet holes was connected to a string (purple solid lines in Fig.~\ref{fig:m-fem}a). The length of these strings was then progressively reduced during the course of the simulation. The pairs of reference points (corresponding to the rivet holes) were connected in a way so that their rotational degrees of freedom were constrained to maintain planarity. For the initiation of the weaving simulation, the first loading step perturbed one vertex within the mesh in the $z$-direction to break the $x$-$y$ plane symmetry. The second loading step disabled this perturbation. A displacement was then applied to the reference points (by shortening the length of the connecting strings) to bring them closer to each other. Eventually, the length of each string (and hence the distance between two reference points within a matching pair) was identically zero, at which point the unit cell was deemed as woven. This procedure is analogous to the sequence of physical weaving performed by hand during the experiments.

Each ribbon was partitioned along the arc length such that the discrete material centerline, $\bm{\gamma}^i_j$, could be extracted from the mesh nodes. As the quadratic mesh edge length was fixed to $\SI{0.75}{\milli\metre}$, the mesh nodes along the arc were approximately equidistant. We follow the same notation introduced in Sec.~\ref{sec:imageprocessing}. An analogous set of mesh nodes along the edge of each ribbon was used to recreate the edge curve (see Fig.~\ref{fig:m-fem}a). These nodes were sorted in a clockwise fashion. The edge-based Cosserat frame was constructed at the midpoint between every two consecutive nodes. The tangent vector $\mathbf{t}^i$ was defined similarly to what we did in the image processing procedure described in Sec.~\ref{subsec:imageprocessing_1}. The binormal vector $\mathbf{b}^i$ was constructed between each mid-point and its closest point on the edge curve. The normal vector $\mathbf{n}^i$ was then obtained from the cross product $\mathbf{t}^i\times \mathbf{b}^i$. With this centerline-based description, the integrated geodesic curvature and the interior angles could be readily obtained in the same way described in Sec.~\ref{subsec:imageprocessing_3}. The total strain energy accumulated at the completion of the weaving process could also be computed directly from the FE simulations.

For $n=\{4, 5, 6, 7, 8\}$, a sampling of the possible parameterizations of $\kappa^*$ assuming $\kappa_1=\pm \kappa_3$, and $\kappa_{1,2,3} \leq \pm 0.5$ with a granularity of $0.1$ led to a total of 536 independent simulations. This enlarged set of parameterizations served to further characterize and gain additional insight into the mechanics of the woven structures. First, the kinematics of the simulations were validated by comparing FEM results with both experiments and the analytical calculations. From these simulation results, we observed that regions of stress concentrations develop near the crossing points during the weaving process. When the weaving is completed, we found that these stress concentrators were distributed more uniformly around the $n$-gon (see Fig.~\ref{fig:m-fem}b).

\begin{figure}[h!]
    \centering
    \includegraphics[width=1\textwidth]{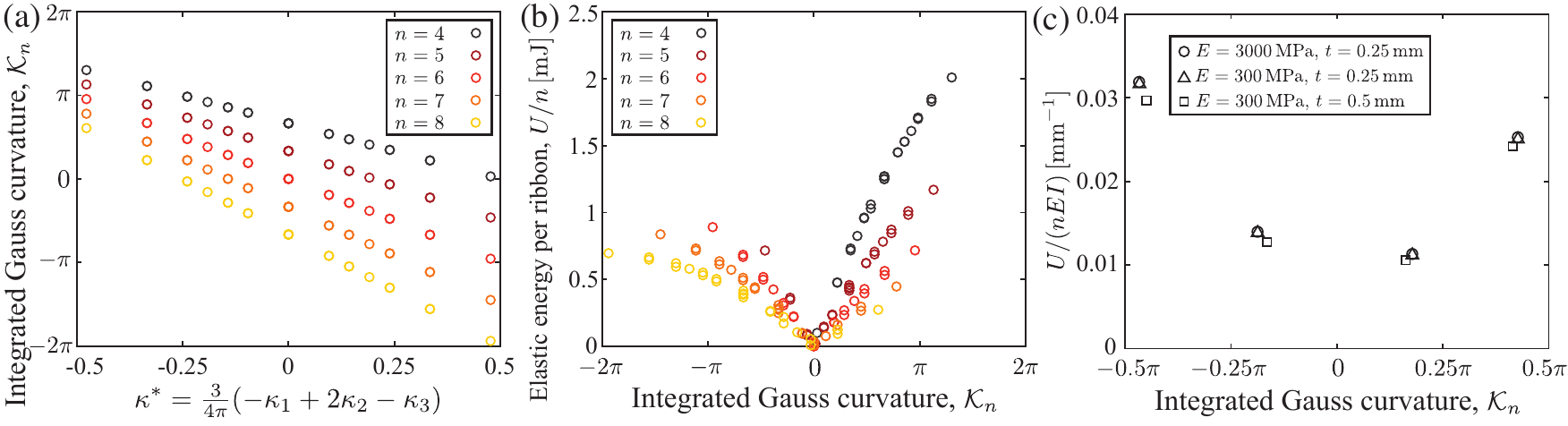}
    \caption{(a) Integrated Gaussian curvature measured from FEM for unit cells woven with $n=\{4,5,\cdots,8\}$, for the different parameterizations of $\kappa^*=\frac{3}{4\pi}(-\kappa_1+2\kappa_2-\kappa_3)$ (same set of simulations as shown in Fig.3b of the main text). (b) Elastic energy per ribbon of the same unit cells plotted in (a). (c) Elastic energy of unit cells normalized by the number of ribbons times bending stiffness of the ribbon, $EI$, as a function of the measured integrated curvature. Three different combinations of $E$ and $I$ are considered (with $E=\{300,\,3000\}\,\si{\mega\pascal}$ and $t=\{0.25,\,0.5 \}\,\si{\milli\metre}$; see legend), for unit cells of $n=6$, $\kappa_2=\{-0.5,-0.2,0.2,0.5\}$ and $\kappa_{1,3}=0$.}
    \label{fig:m-fem_data}
\end{figure}

In Fig.~\ref{fig:m-fem_data}(a), we plot numerical results for the relationship between the integrated curvature of the unit cell, $\mathcal{K}_n$, and its injected in-plane curvature, $\kappa^*$. We observe that multiple parametrizations of the unit cell (multiple values of $\kappa^*$) can lead to the same value of $\mathcal{K}_n$; \textit{e.g.}, an integrated Gauss curvature of $\mathcal{K}_n=\pi/3$ can be achieved with, but not limited to, $(n,\kappa^*)=(4,0.25),(5,0),(6,-0.167), (7,-0.285), (8,-0.375)$. As shown analytically in Eq.~(6) of the main text, to increase the value of $\mathcal{K}_n$, one can either increase the number of ribbons $n$ or decrease the value of $\kappa^*$.

The elastic energy of each simulated system once fully woven is plotted Fig.~\ref{fig:m-fem_data}(b) for each parametrization set of the unit cell. While a desired $\mathcal{K}_n$ can be achieved using multiple parameterizations, Fig.~\ref{fig:m-fem_data}(b) shows that the elastic energy stored by each ribbon between these parameterizations can vary significantly. 
Therefore, in addition to satisfying topological constraints (Euler characteristics), the elastic energy of the woven ribbons can inform the designer regarding which parameter set to use when weaving towards a target shape; \textit{i.e.}, one would choose the parametrization with the lowest energy that still results in the desired $\mathcal{K}_n$. This observation begets a formal optimization, outside the scope of this paper, where $\kappa_{1,2,3}$ are tuned globally such that the target surface is attained while minimizing the total strain energy. We hope that future work will address this (far more) complex optimization problem.

To further justify that the shape of the weaves is controlled by $\kappa^*$, independently of the shape and rigidity of the ribbons, we conducted two additional sets of simulations for $n=6$, $\kappa_2=\{-0.5,-0.2,0.2,0.5\}$ and $\kappa_{1,3}=0$. For the first set of four simulations, Young's modulus of the ribbons was set to $E=3000\,\text{MPa}$ and the thickness was $t=\SI{0.25}{\milli\metre}$. Then, the bending modulus of the ribbons is varied by changing the material properties by taking $E=300\,\text{MPa}$, and the moment of inertia by considering two thicknesses values, $t=\SI{0.25}{\milli\metre}$ and $t=\SI{0.50}{\milli\metre}$. In Fig.~\ref{fig:m-fem_data}(c), we plot the elastic energy of the unit cells ($U$) normalized by the bending stiffness of the ribbons, $EI = \frac{Ewt^3}{12(1-\nu^2)}$, and the number of ribbons, $n$, as a function of the measured integrated curvature $\mathcal{K}_6$. The three sets of simulations collapse onto the same values. The collapse along the horizontal axis, $\mathcal{K}_6$, shows that the shape of weaves is independent of both Young's modulus and the cross-sectional profile of the ribbons. The collapse along the vertical axis, $U/(nEI)$, shows that most of the energy of the ribbons is stored into bending and not into stretching nor twisting energies. 

\subsection{Developability of exterior triangles in the unit cells}
\label{subsec:triangles}

In Fig.~2(d) of the main text where we plotted data from a series of experiments and FEM simulations, we showed that the average interior angle of the \textit{n}-gon is approximately $\langle \theta^\circ \rangle \approx 2\pi/3$. We interpreted this observation geometrically as a statement that the exterior triangles of the unit cell remain nearly developable. This developable triangle assumption was then extended to the case of unit cells with initially curved ribbons, from which we derived the expression for the integrated curvature of unit cells; Eq.~(4) of the main text. In this section, we quantify the developability of adjacent triangles in the unit cells.

\begin{figure}[h!]
    \centering
    \includegraphics[width=0.5\textwidth]{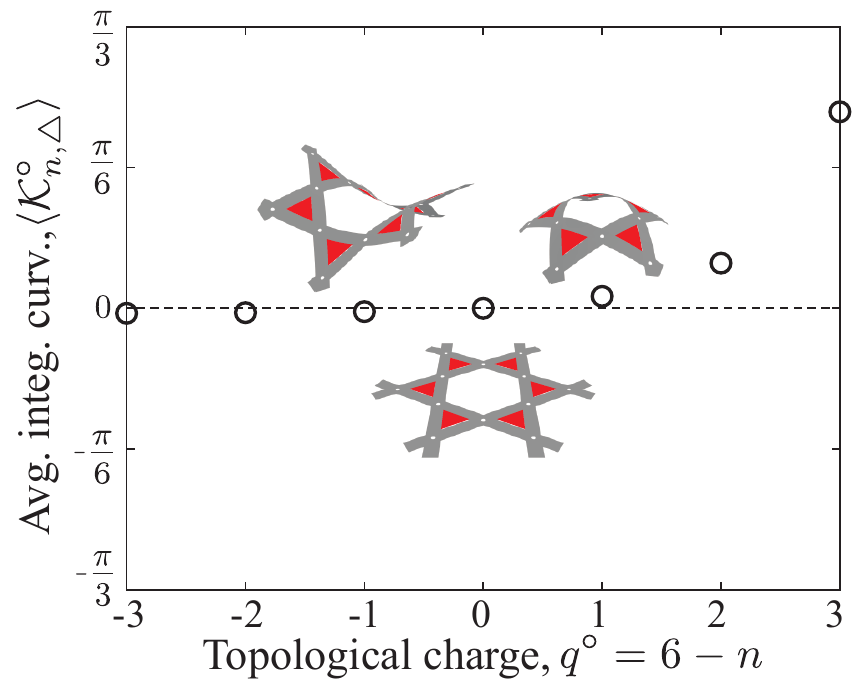}
    \caption{Average integrated curvatures of the exterior triangles (red shaded regions in the inset) in unit cells, $\langle \mathcal{K}^{\circ}_{n,\triangle}(n) \rangle$, \textit{vs}. $n$, measured from experiments. The triangles remain nearly developable, \textit{i.e.}, $\langle \mathcal{K}^{\circ}_{n,\triangle}(n) \rangle = 0$, for cases $q \le 1$, or equivalently $n \ge 5$. These data were obtained from experiments (using $\mu$CT tomography), for the same samples studied in Fig.~2 of the main text, but now focusing on the external triangles, instead of the central $n$-gon. }
    \label{fig:triangles_curvature}
\end{figure}

We have computed the averaged integrated Gaussian curvature of the triangles in unit cells, $\langle \mathcal{K}^{\circ}_{n,\triangle}(n) \rangle$, with different values of \textit{n} and plot the results in Fig.~\ref{fig:triangles_curvature}. Following the convention used in the main text, the superscript $(\cdot)^\circ$ is used to indicate quantities associated with initially straight ribbons. The integrated Gaussian curvature of each triangle, $\mathcal{K}^{i}_{n,\triangle}(n)~(1 \le i \le n)$, is obtained by measuring the interior angles of the triangles and applying the Gauss-Bonnet theorem; Eq.~(1) of the main text. Due to the rotational symmetry of the unit cell, $\mathcal{K}^{i}_{n,\triangle}(n)$ does not vary for different values of $i$, thus it is sufficient to present their average, 
\begin{equation}
\langle \mathcal{K}^{\circ}_{n,\triangle}(n) \rangle = \frac{1}{n}\sum_{i=1}^{n} \mathcal{K}^{i}_{n,\triangle}(n).
\end{equation}
For the cases with $n \ge 5$ (\textit{i.e.}, $ q^\circ \le 1$), we find that $\langle \mathcal{K}^{\circ}_{n,\triangle}(n) \rangle \simeq 0$ and the triangles in the unit cell remain nearly developable. However, $n = \{3, 4\}$ (\textit{i.e.}, $q^\circ = \{2, 3\}$) appear to be extreme cases, for which the developability condition appears to break down as  $q^\circ$ increases beyond $1$. While this is a very interesting result that deserves further attention, we have not yet been able to develop a predictive argument to rationalize it. We hope that the present findings, including the limiting cases of the developability condition of the external triangles, will instigate additional future research to investigate their role in dictating the mechanics of triaxial weaves.

\subsection{Bending energy of the unit cells}
In Section~\ref{subsec:triangles} above, we observed that the exterior triangles remain nearly developable, especially for $n \ge 5$, and the Gaussian curvature concentrates at the \textit{n}-gon. If the \textit{n}-gon were not constrained by the adjacent triangles, both the \textit{n}-gon and the triangles would remain flat and, respectively, have their interior angles equal to $180^\circ-360^\circ/n$ and $60^\circ$. However, since the interior angles of the triangles and the \textit{n}-gon must add up to $180^\circ$ (the two angles are supplementary to each other); there is a competition between the angles of the adjacent triangles and the \textit{n}-gon. We observe that the adjacent triangles always `win' in this competition and remain developable in the woven configuration. Due to this strong geometric constraint imposed by the exterior triangles, the average interior angle of the \textit{n}-gon is approximately $120^\circ$, as evidenced by the data in Fig.~2(d) of the main text. (It is important to note that the case with $n=3$ appears as an exception where some deviations are observed. This limiting case will be addressed below in more detail.) Consequently, an angular deficit (or surplus) is introduced into the \textit{n}-gon, leading to the negative (or positive) integrated Gaussian curvature of the \textit{n}-gon, depending on the value of $n$, while the adjacent triangles remain with nearly zero integrated curvature throughout.

To provide a tentative rationale for these findings described in the previous paragraph, we construct an energy-based argument based on the grounds that it is (energetically) more costly to induce surface curvature on the triangles than on the $n$-gon. While it is beyond the scope of the present study to resolve the elastic energy of full-scale weaves with multiple topological defects, we believe it is informative to consider the isolated unit cells and investigate the effect of $n$ on its elastic energy. We will estimate the shape and energy of the $n$-gons using two simplified models: each based on a `\textit{spherical}’ and a `\textit{conical}’ geometry, respectively, for the underlying geometry. Eventually, the results from both the `\textit{spherical}’ and the `\textit{conical}’ models will be compared to the numerical (FEM) results.

We start by considering a `\textit{spherical}' model for the underlying geometry, making the following simplifying assumptions on the geometry of the unit cell:
\begin{itemize}
    \item The bending energy of the unit cell is restricted to the ribbon’s normal direction.
    \item The spanning surface of the \textit{n}-gon, over which the Gauss-Bonnet theorem will be applied, is approximated to a sphere (for positively curved surfaces). The edges of the \textit{n}-gon are then regarded as great circles of this sphere.
    \item The part of the ribbons outside of the \textit{n}-gon edges does not store elastic energy, thereby remaining straight. 
\end{itemize}

Given the above assumptions for the \textit{spherical} model, one can propose a direct link between the integrated curvature of the \textit{n}-gon, $\mathcal{K}_n$, and the normal curvature of the ribbons. With a spanning area approximated by a sphere, the integrated Gaussian curvature of the \textit{n}-gon is $\mathcal{K}_n=A/R^2$, where $R$ is the radius of the sphere and $A$ is the area of the spherical cap spanned by the \textit{n}-gon. Since the ribbons were approximated as geodesics, their centerlines can be regarded as great circles of the sphere, thus with a normal curvature of $1/R$. The energy of the \textit{n}-gons can then be computed as $U=n\frac{EI}{2}\frac{\ell}{R^2}$ ($EI$ is bending stiffness of the ribbons in the normal direction, and $\ell$ the length of a segment). Then, we roughly approximate the area $A=n\frac{\sqrt{3}}{4}\ell^2$ simply by the area in the flat case ($n = 6$). Using the expression of $\mathcal{K}_n$ derived in the main text, Eq. (3), we obtain an expression for the energy of the \textit{n}-gon as a function of both the geometry $(\ell,\kappa^*)$ and the number of ribbons $n$:
\begin{equation}
    \frac{U\ell}{EI}=\frac{2\pi}{3 \sqrt{3}}\left[{6-n(1+\kappa^*)}\right].
\end{equation}

\begin{figure}[h!]
    \centering
    \includegraphics[width=1\textwidth]{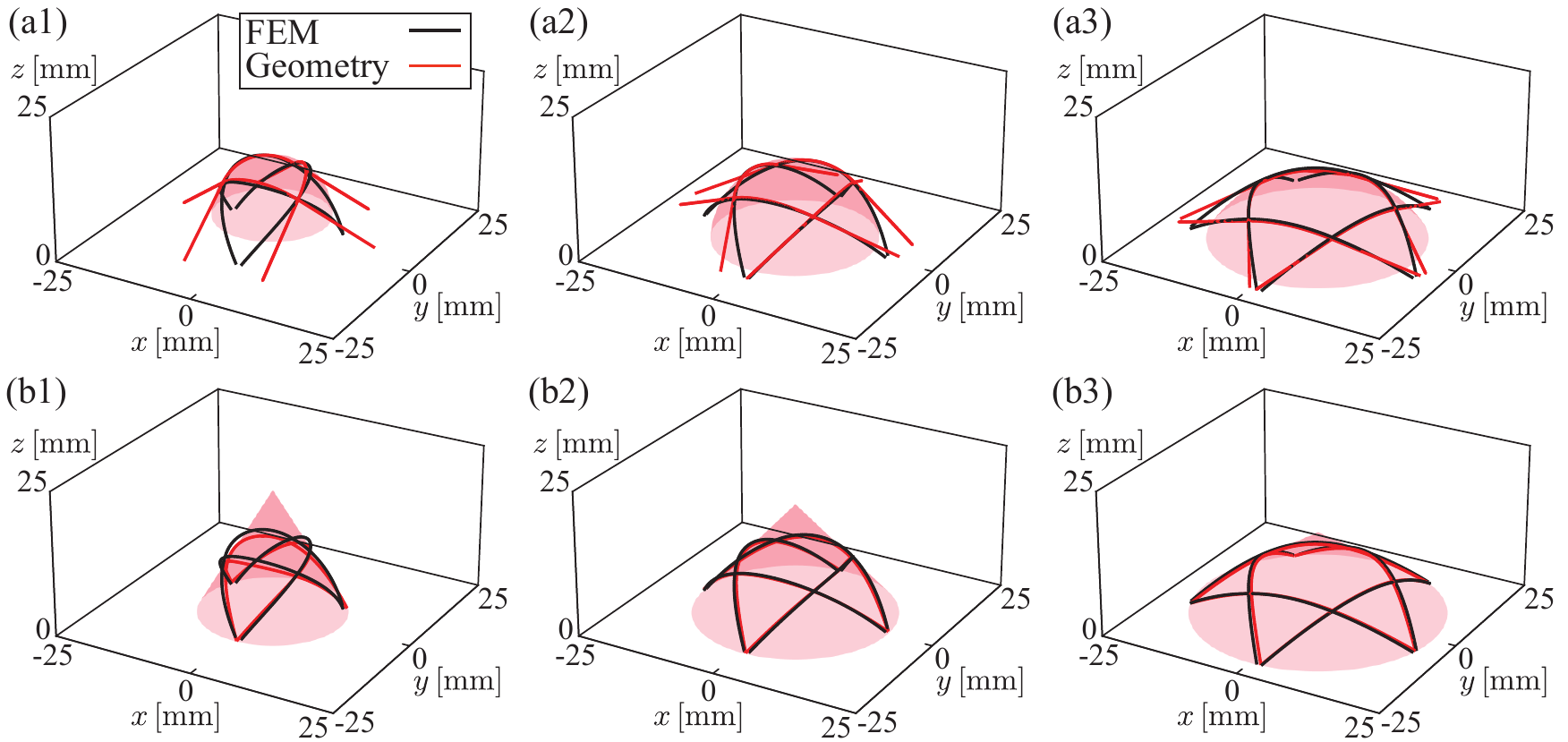}
    \caption{Visualization of unit cells with $n = \{3, 4, 5\}$, with two different spanning surface approximations: (a) the \textit{spherical} model, and (b) the \textit{conical} model. Black solid lines are from FEM simulations and red solid lines are the computed geodesics laying on the spanning surfaces.}
    \label{fig:geodesic_shape}
\end{figure}

\begin{figure}[h!]
    \centering
    \includegraphics[width=0.5\textwidth]{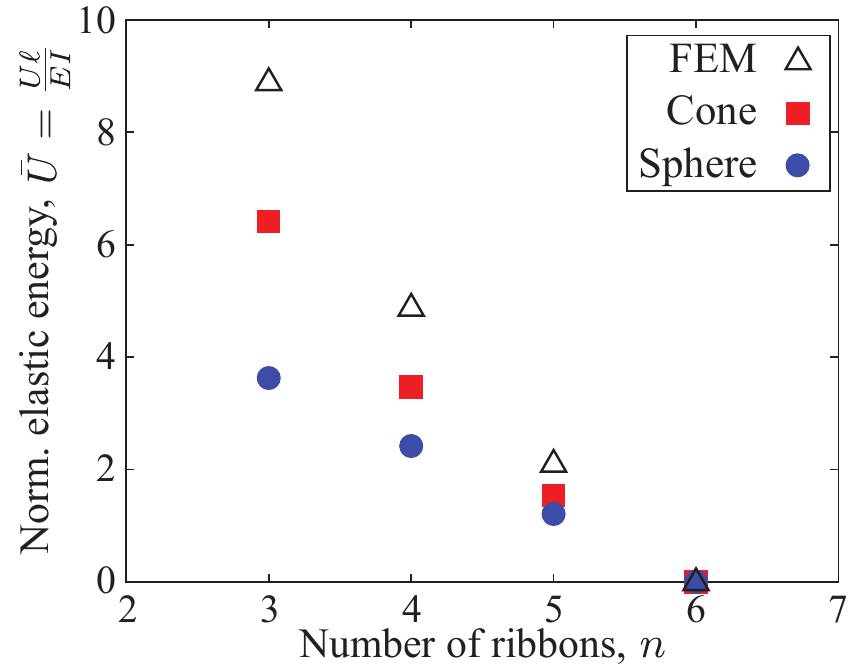}
    \caption{Normalized elastic energy of the unit cells, $\bar{U}=U\ell/EI$, vs. the number of ribbons, $n$. The cone model shows greater agreement with the FEM measurement.}
    \label{fig:geodesic_energy}
\end{figure}

Within the framework of the \textit{spherical} model presented above, the energy of the \textit{n}-gons is, indeed, a decreasing function of \textit{n}, with the deformation of the triangles being more energetically costly than other polygons. This finding is consistent with our numerical (FEM) observations, as will be evidenced below when discussing the results in Fig.~\ref{fig:geodesic_energy}. From this \textit{spherical} model, we can also derive the corresponding trajectories of the ribbons from geometrical arguments. The central part of the ribbons lies on great circles of the sphere, connected at their ends at the pins. By symmetry, each edge must span an azimuthal angle $2\pi/n$, with arc length $\ell$. Making use of the relation between the arc length and the starting-ending azimuthal and polar angle for great circles of a sphere (haversine formula), we can then deduce that the polar angle of the pins $\lambda$ must satisfy $\cos(\ell/R)=\cos^2\lambda+\sin^2\lambda\cos(2\pi/n)$. The trajectories of the great circles corresponding to the \textit{n}-gon edges can then be computed. The part of the ribbons outside of the \textit{n}-gons are simply drawn straight; see Fig.~\ref{fig:geodesic_shape}(a). 

Next, we consider an equivalent \textit{conical} model. Whereas the above \textit{spherical} model assumed that the \textit{n}-gon spans a sphere (while the remaining external edges remain undeformed), the \textit{conical} model will assume that the whole unit cell fits onto a \textit{conical} surface. Indeed, if the straight ribbons are only deformed by normal bending, they should all lay on a same conical (developable) surface, with their trajectory resting on geodesics of this surface. Under the assumption that the spanning surface of the \textit{n}-gon is a cone, one can now directly relate the integrated curvature of the \textit{n}-gon $\mathcal{K}_n^o$ to the cone geometry; $\mathcal{K}_n^o=2\pi(1-\sin(\gamma))$ where $\gamma$ is the cone’s opening angle $(0<\gamma<\pi/2)$. Then, using Eq.~(2) of the main text, a simple expression can be obtained for this opening angle as a function of $n$:
\begin{equation}
    \sin(\gamma)=\frac{n}{6}.
    \label{eq:cone_angle}
\end{equation}
It is well established \cite{Stoker:1969vt} that the geodesics on a cone with a prescribed opening angle $2\varphi_0$ (starting and ending at a radius $\rho_0$ of the cone) can be expressed as:
\begin{equation}
    \rho(\phi)=\rho_0\left[{\frac{\cos\left({\sin(\gamma)\varphi_0}\right)}{\cos\left({\sin(\gamma)\varphi}\right)}}\right],
    \label{eq:cone_geo}
\end{equation}
where $\varphi\in[-\varphi_0, \varphi_0]$ is the polar angle around the axis of the cone and $\rho$ is the radial coordinate of the cone. With this result, it is then possible to obtain different geodesics depending on the values of $\rho_0$ and $\varphi_0$. In order to find the geodesics of this cone corresponding to the trajectories of the ribbons, the two following conditions on the geodesics have to be respected: %
\begin{enumerate}[(i)]
    \item Given the symmetry of the \textit{n}-gon, the intersections with the base of the cone must be separated by an angle $4\pi/n$, thereby imposing $\varphi_0=2\pi/n$.
    \item The total arc length of these geodesics must be equal to $3\ell$; \textit{i.e.}, the total arc length of a ribbon. This condition defines uniquely the radius of the cone’s base $\rho_0$.
\end{enumerate}
Therefore, without any adjustable parameters, we obtain the expression of the geodesics corresponding to the trajectory of a ribbon. The other $n-1$ ribbons' trajectories can be readily determined by rotations of $2\pi/n$ around the cone axis. 

In Fig.~\ref{fig:geodesic_shape}(b1-3), we compare the prediction of the centerlines from the geodesics of the cone (Eq.~\ref{eq:cone_geo}; thin red lines) with the FEM-computed centerlines (thick black lines), for unit cells with $n=\{3,4,5\}$. We focus on these three cases so as to give particular attention to the limiting cases mentioned above where the developability of the external triangles is debatable. In each of the plots in Fig.~\ref{fig:geodesic_shape}(b1-3), we also represent the conical surfaces (in red) with opening angles that were computed from Eq.~\ref{eq:cone_angle}. The \textit{conical} model captures remarkably well the full trajectories of the ribbons for the cases $n=\{4,5\}$. However, for the extreme case with $n=3$, we still observe some deviations, highlighting that this is a far from trivial case, calling for an even more sophisticated analysis, which we hope will be addressed in future work. For comparison purposes, in Fig.~\ref{fig:geodesic_shape}(a1-3), we also present the predictions from the \textit{spherical} model (thin red lines) discussed above. In this simpler model, since we assumed that only the \textit{n}-gon edges were embedded on a sphere, by construction, the \textit{spherical} model cannot accurately capture the trajectories of the surrounding triangles. Overall, we find that the \textit{conical} model proves a more satisfactory description of the unit cells than the \textit{spherical} model.

The trajectory of the ribbon centerlines obtained from either the \textit{spherical} or the \textit{conical} models can now be used to estimate the elastic energy of the unit cells by numerically integrating the normal bending energy along the centerline trajectories. In Fig.~\ref{fig:geodesic_energy}, we plot the estimated normalized elastic energy, $\bar{U}=U\ell/EI$, as a function of the number of ribbons, $n$. Similar to the experiments, the segment length of the ribbons was fixed to $\ell = \SI{15}{\milli\metre}$. Overall, we find that the \textit{conical} model outperforms the \textit{spherical} model, although both underestimate the elastic energy compared to the FEM results. Still, both models predict the general trend that the energy of the unit-cells decreases with $n$. The present results based on a geometric reasoning, combined with the FEM simulations, provide a tentative rationale, even if only qualitative, for the statement that it is energetically more costly to induce Gaussian curvature on the triangles than on the $n$-gons, which is employed in the general argument of the main text.

In conclusion to this section, we summarize the key aspects and provide a comparative discussion of the \textit{spherical} and the \textit{conical} models:
\begin{itemize}
    
    \item The \textit{conical} model accurately captures the full shape of the ribbons, even if still only approximately for $n=3$. The drawback of this model is that it can only be applied for the geodesics of the cones; \textit{i.e.}, when $\kappa^*=0$.
    
    \item The \textit{spherical} model yields analytical expressions and is applicable for $\kappa^* \neq 0$, but is only valid for very shallow shells, for which $n(1+\kappa^*)\approx6$.
    
    \item Both models provide satisfactory predictions for the shape of the ribbons along the edges of the \textit{n}-gon, which is remarkable given the strong geometric nature of the argument.
    
    \item Both models provide better than just order-of-magnitude estimates for the stored elastic energy and decreasing values with \textit{n}.
    
    \item Both models underestimate the total elastic energy of highly curved unit cells, even if the conical model is more accurate. The observed mismatch, especially for $n = {3, 4}$, suggests that highly curved weaves show other modes of deformations than normal bending of the ribbons. It is possible that the nontrivial mechanics of ribbons (in between a rod and a plate) plays a role and would need to be included in more advanced descriptions.
    
    \item Both models are restricted to positively curved weaves. More sophisticated geometric arguments, perhaps using pseudoconic projections, may need to be invoked to describe the negatively shaped cells.
    
\end{itemize}

\section{Design of (simple) non-spherical weaves}
In this section, we detail the design procedure that we developed to weave simple non-spherical shapes by using piecewise circular ribbons. We shall focus specifically on ellipsoidal and toroidal weaves, noting that more complex geometries and topologies would require a level of numerical optimization beyond the scope of the present work.

\begin{figure}[h!]
    \centering
    \includegraphics[width=0.9\columnwidth]{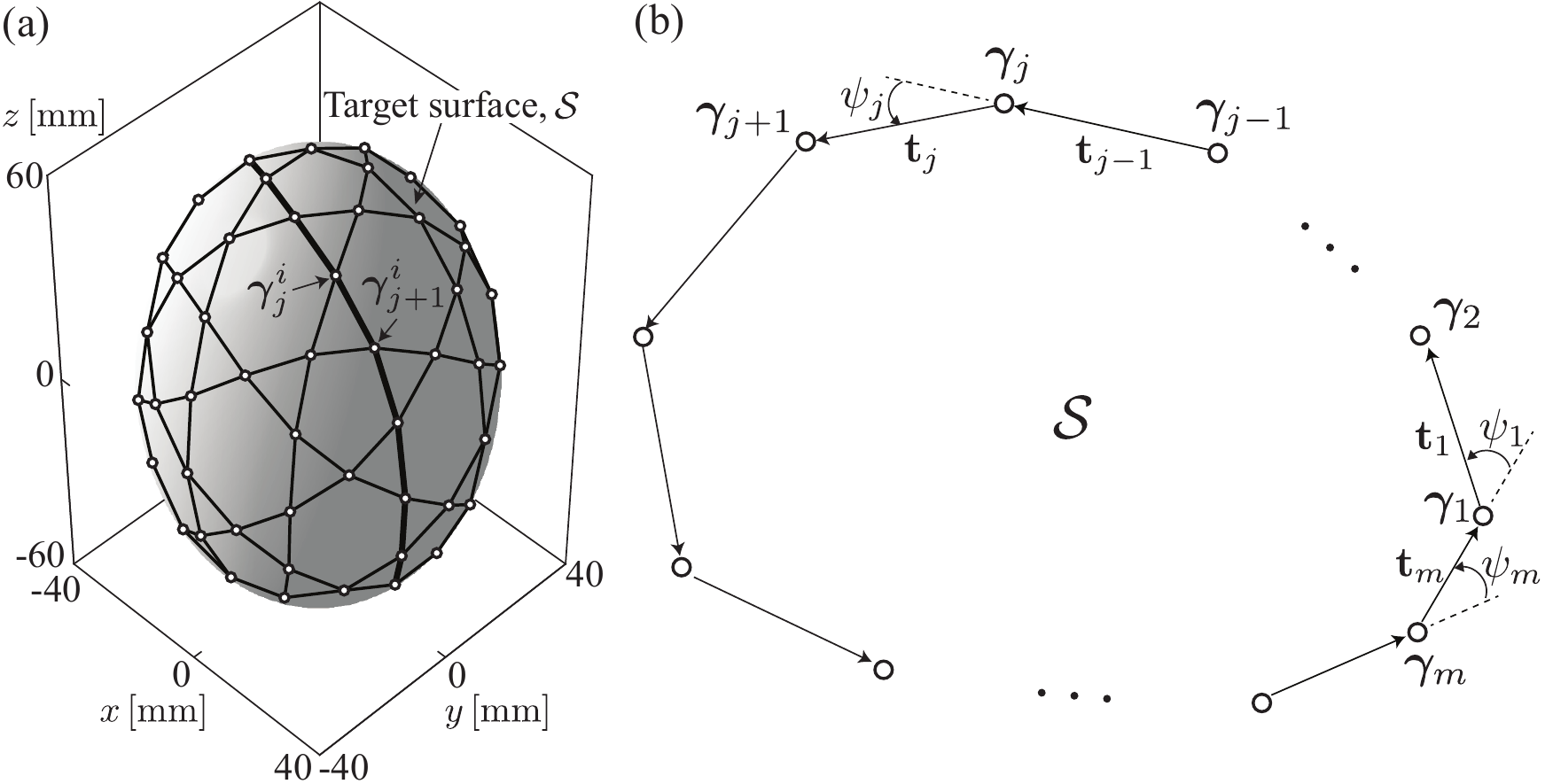}
    \caption{(a)~Topology of the ellipsoidal weave of a polar radius $a=\SI{60}{\milli\metre}$, and an equatorial radius $b=\SI{40}{\milli\metre}$. The curvature and arc length of the $j$-th arc of the $i$-th ribbon, $(\kappa_j,\,\ell_j)$, is obtained by calculating the geodesic turning angle at the adjacent crossing points, $\bm{\gamma}_{j}^i$ and $\bm{\gamma}_{j+1}^i$ (crossing points are denoted as white circles). (b)~The discrete curve $\{\bm{\gamma}_j\}$ is embedded on the target surface $\mathcal{S}$. The curve consists of $m$ vertices and edges, which we denote as $\bm{\gamma}_j$ and $\mathbf{t}_j$, respectively. The discrete integrated geodesic curvature that is associated to each vertex is denoted by $\psi_j$.}
    \label{fig:ellipsoid_topology}
\end{figure}

Fig.~\ref{fig:ellipsoid_topology}(a) shows the required inputs for our design procedure: the surface $\mathcal{S}$ of the target shape (taking the example of an ellipsoid) and a graph on $\mathcal{S}$ representing the topological layout of the weave. Each vertex of the graph represents the crossing points between two ribbons, and each edge represents the connectivity of the crossing points. Note that the edges of the graph do not represent the actual ribbons. From the topological layout in Fig.~\ref{fig:ellipsoid_topology}(a), we label the location of each vertex as $\bm{\gamma}^i_j~(1 \le j \le m^i)$, where the superscript $i$ is the index of the ribbons, the subscript $j$ is the index of the crossings comprising the $i$-th ribbon of the weave, and $m^i$ is the number of the crossing points on the $i$-th ribbon. In short, $\{\bm{\gamma}_j^i\}~(1 \le j \le m^i)$ is an ordered array of the crossing points representing the $i$-th ribbon of the weave.

\subsection{Computing the shape of the piecewise circular ribbons}
\label{subsec:inversedesign}
Our goal is to compute the undeformed geometry of the $i$-th ribbon from the position of its crossings points, $\{\bm{\gamma}^i_j\}$. Hereafter, for the sake of brevity, we omit the superscript $i$ (the ribbon index). In Fig~\ref{fig:ellipsoid_topology}(b), we sketch a typical example of $\{\bm{\gamma}_j\}$, onto which we superpose a representation of the vertices and edges. We regard the set of crossing points $\{\bm{\gamma}_j\}$ as a discrete curve embedded on $\mathcal{S}$, representing the shape of the woven ribbon. We denote the $j$-th edge of the discrete curve, $\mathbf{t}_j$, as a vector connecting $\bm{\gamma}_j$ and $\bm{\gamma}_{j+1}$; $\mathbf{t}_j = \bm{\gamma}_{j+1} - \bm{\gamma}_{j}$. Since the ribbon is closed when woven, the number of edges of the discrete curve is $m$. Accordingly, we define $\mathbf{t}_m = \bm{\gamma}_{1} - \bm{\gamma}_{m}$.

Next, we choose to approximate the initial shape of a ribbon as a piecewise-circular curve comprising $m$ segments. Denoting the length and curvature of the $j$-th segment of the undeformed ribbons by $\ell_j$ and $k_j$, respectively, the design problem is reduced to estimating $(\ell_j,\,k_j)~(1 \le j \le m)$ from $\{\bm{\gamma}_j\}$. To do so, first, we compute $\ell_j$ by taking the length of the $j$-th edge: 
\begin{equation}
    \ell_j = |\mathbf{t}_j|.
    \label{eq:segmentlength}
\end{equation}
As a second step, we need to obtain $k_j$. We compute the discrete geodesic curvature of the curve at each vertex, $\psi_{j}$, by quantifying the rotation of the adjacent edges, $(\mathbf{t}_{j-1}$ and $\mathbf{t}_j)$:
\begin{equation}
    \psi_{j} = \bigg( \frac{2 \mathbf{t}^i_{j-1} \times \mathbf{t}^i_{j}}{ |\mathbf{t}^i_{j-1}| |\mathbf{t}^i_{j}| + \mathbf{t}^i_{j-1} \cdot \mathbf{t}^i_{j} } \bigg) \cdot \mathbf{N}_j(\bm{\gamma}_j)
    \label{eq:psi}
\end{equation}
where $\mathbf{N}(\bm{\gamma}_j)$ is the normal of the target surface $\mathcal{S}$ at $\bm{\gamma}_j$~\cite{bobenko2008discrete, bergou2008DER}.
Note that, in this discrete setting, $\psi_{j}$ is defined at each vertex. By contrast, the piecewise-constant curvature of the undeformed ribbon, $k_j$, is associated with each edge. Therefore, we define the integrated geodesic curvature of the $j$-th segment of the undeformed ribbon, $\kappa_j = k_j \ell_j$, as the average of the geodesic curvature of the vertices adjacent to the $j$-th edge:
\begin{equation}
    \kappa_j = \frac{\psi_{j} + \psi_{j+1}}{2}
    \label{eq:inversedesign_kappa}
\end{equation}
where $\psi_{j}$ was defined in Eq.~\eqref{eq:psi}.

For a closed discrete curve, an example of which is depicted in Fig.~\ref{fig:ellipsoid_topology}(b), its last ($\bm{\gamma}_m$) and first ($\bm{\gamma}_1$) points are connected. Hence, we note that, for the cases when $j=1$ and $m$, we need exceptions to the definition of $\psi_j$ in Eq.~\eqref{eq:psi}) and $\kappa_j$ (Eq.~\eqref{eq:inversedesign_kappa}). When $j=1$, $\psi_1$ is computed from $\mathbf{t}_m$ and $\mathbf{t}_1$ ($\mathbf{t}_0$ is not defined). When $j=m$, we define $\kappa_m=(\psi_{m} + \psi_{1})/{2}$, since the last edge ($j=m$) is associated to the pair of points $\bm{\gamma}_m$ and $\bm{\gamma}_1$. By doing so, Eqs.~\eqref{eq:psi} and \eqref{eq:inversedesign_kappa} are well defined in the full range $1 \le j \le m$.

Our design framework requires that both the location of the vertices and the topology of the weave, $\{\bm{\gamma}^i_j\}$, be pre-determined as inputs. The systematic generation of the topological layout for arbitrary target shapes is beyond the scope of our study. Still, we can design simple yet canonical shapes, such as ellipsoids and tori, whose topological layout can be readily obtained due to their symmetries. In Sec.~\ref{subsec:weave_inversedesign_ellipsoid} and Sec.~\ref{subsec:weave_inversedesign_torus}, we will design the initial shape of the ribbons to construct ellipsoidal and toroidal weaves, based on Eqs.~\eqref{eq:segmentlength} and \eqref{eq:inversedesign_kappa}.

\subsection{Ellipsoidal weave}
\label{subsec:weave_inversedesign_ellipsoid}

\begin{figure}[h!]
    \centering
    \includegraphics[width=0.8\columnwidth]{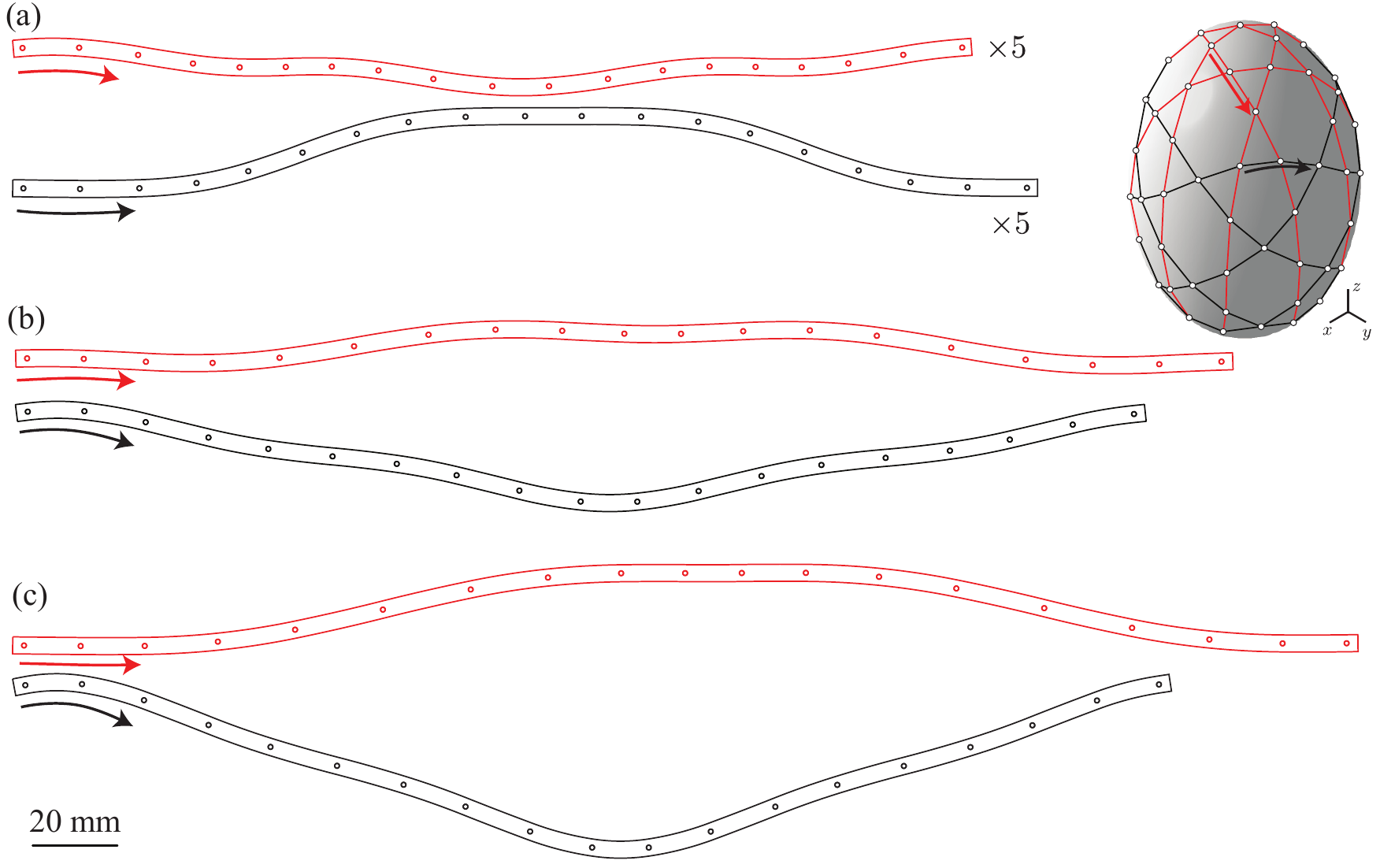}
    \caption{Planar shapes of the ribbons required to weave the three ellipsoidal weaves presented in Fig.~4a of the main text: (a)~$a=\SI{30}{\milli\metre},\,b=\SI{40}{\milli\metre}$, (b)~$a=\SI{50}{\milli\metre},\,b=\SI{40}{\milli\metre}$, and (c)~$a=\SI{60}{\milli\metre},\,b=\SI{40}{\milli\metre}$. Note that the ribbons form closed curves when woven (the first and the last crossings meet in their final configuration). The weaves consist of five ribbons that pass the pentagons at the north and south poles (red ribbons) and five black ribbons that do not (black ribbons). Given the common scale bar, these ribbons could be printed and woven by the interested reader. The blueprint of these ribbons is provided as vector files in the Supplementary Material~\cite[]{SupInf_PRL}.}
    \label{fig:ellipsoid_blueprint}
\end{figure}

As a first example, we consider an ellipsoidal weave with an equatorial radius, $b$, and a polar radius, $a$, such that the aspect ratio is $a/b$. The target shape of this ellipsoid, along with the underlying topological layout, were presented above, in Fig.~\ref{fig:ellipsoid_topology}. As a starting point of the topological layout, we obtained the coordinates of the vertices from the coordinates of the \textit{Rectified truncated icosahedron}~\cite{atI1, atI2}. Hence, the ellipsoidal weave will contain 12 pentagons, 20 hexagons, and 60 triangles, just as the spherical weave in Fig.~3 of the main text. Then, we performed a linear scaling of the layout by a factor $a$ along the $x$ and $y$ axes and a factor $b$ along the $z$ axis so as to match the layout as closely as possible to an ellipsoid with the specified equatorial and polar radius. In Figs.~\ref{fig:ellipsoid_blueprint}, we present the shape of undeformed ribbons of our three ellipsoidal weaves with an equatorial radius, $b=\SI{40}{\milli\metre}$, and polar radii, $a\,[\si{\milli\metre}]=\{30,\,50,\,60\}$, corresponding to the examples presented in Fig.~4a of the main text. The complete layout of the ribbons for the ellipsoidal weaves, including both the shape of the ribbons and their connectivity, is provided as a vectorized blueprint in the Supplementary Material~\cite[]{SupInf_PRL}. These ribbons can printed (or laser cut) and woven by the interested reader. The geometry of the ribbons at their rest configurations were computed from Eqs.~\eqref{eq:segmentlength} and \eqref{eq:inversedesign_kappa}. In Fig.~\ref{fig:ellipsoid_blueprint}, the crossing points of the ribbons are denoted as circles. The first and last circles in each ribbon are supposed to overlap in their woven configuration to form closed loops. 

Due to the axisymmetric shape of the ellipsoid, there are only two distinct types of ribbons (color-coded as red and black in Fig.~\ref{fig:ellipsoid_blueprint}) among the ten ribbons required for the ellipsoidal weave. The red and black arrows in Fig.~\ref{fig:ellipsoid_blueprint} and the inset of Fig.~\ref{fig:ellipsoid_blueprint} are drawn to illustrate the placement of each ribbon in its planar and woven states. The five ribbons of the first type (red ribbons in Fig.~\ref{fig:ellipsoid_blueprint}) pass the pentagons at the north and south poles, and the other five ribbons of the second type (black ribbons in Fig.~\ref{fig:ellipsoid_blueprint}) do not. For the given set of parameters describing the geometry of the ellipsoidal target surface, our algorithm yields ribbon designs that are highly irregular and curved. It would have been unfeasible to arrive at these designs by an iterative \textit{trial and error} approach.

\subsection{Toroidal weave}
\label{subsec:weave_inversedesign_torus}

\begin{figure}[h!]
    \centering
    \includegraphics[width=\columnwidth]{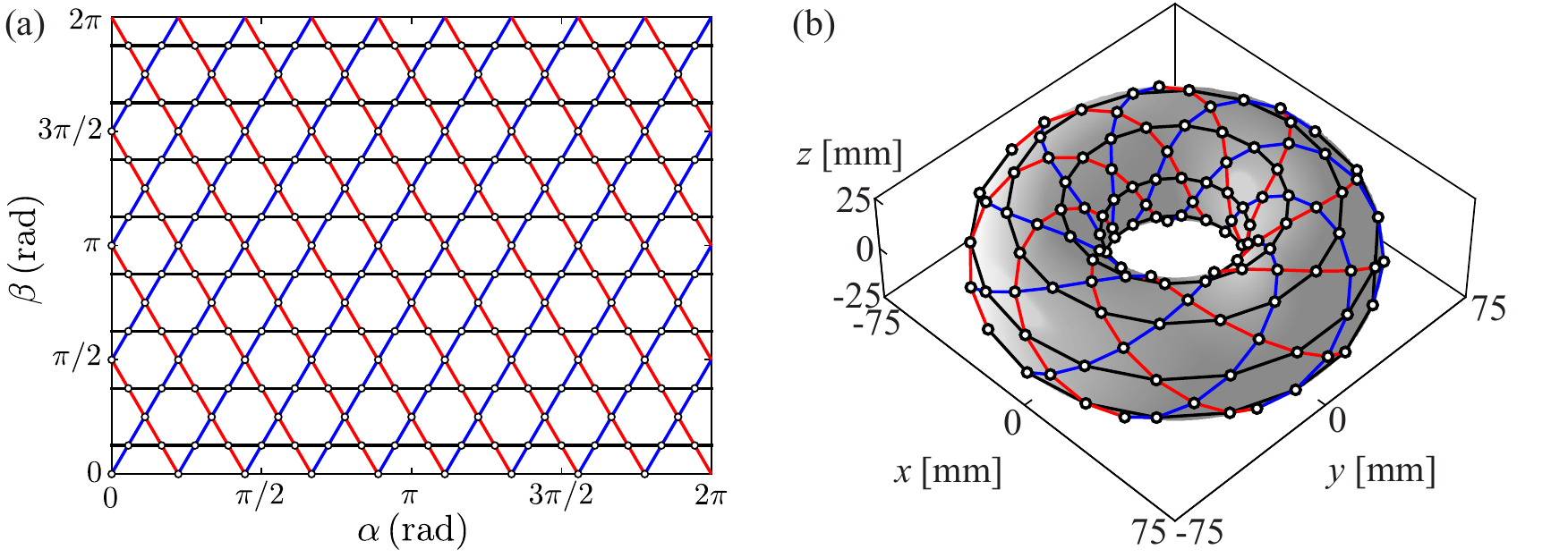}
    \caption{Topology of the toroidal weave with $r_\text{o}=\SI{105}{\milli\metre}$, $r_\text{i}=\SI{35}{\milli\metre}$, $n_\alpha = 9$, $n_\beta = 4$. (a)~The triaxial pattern on the toroidal-poloidal domain ($\alpha$-$\beta$). (b)~The mapping of the triaxial pattern in (a), using Eq.~(\ref{eq:torus}), yields the topology used for our toroidal weave design.}
    \label{fig:torus_topology}
\end{figure}

As a second example of a simple non-spherical weave, we design a smooth toroidal weave with piecewise-circular ribbons. The inner radius of the torus is $r_\text{i}$ and the outer radius is $r_\text{o}$. We only consider the case with $r_\text{i} > 0$, where the target surface does not self-intersect. The idea is to generate the topology of the toroidal weave on a rectangular domain and then map it into a torus by considering the classical parametric expression for the torus~\cite{Stoker:1969vt}:
\begin{equation}
    \begin{aligned} 
        x &= \bigg( \frac{(r_\text{o} + r_\text{i})}{2} + \frac{(r_\text{o} - r_\text{i})}{2} \cos \beta \bigg) \cos \alpha \\ 
        y &= \bigg( \frac{(r_\text{o} + r_\text{i})}{2} + \frac{(r_\text{o} - r_\text{i})}{2} \cos \beta \bigg) \sin \alpha \\ 
        z &= \frac{(r_\text{o} - r_\text{i})}{2} \sin \beta.
    \end{aligned}
\label{eq:torus}
\end{equation}
We defined the toroidal angle coordinate as $\alpha$ (along the direction of the small circular ring around the surface) and the poloidal angle coordinate as $\beta$ (along the direction of the large circular ring around the torus, encircling the central void). In Fig.~\ref{fig:torus_topology}(a), we present the rectangular domain $(\alpha,\beta) \in [0,2\pi) \times [0,2\pi)$, where we draw the regular periodic hexagonal (triaxial) pattern that repeats along the $\alpha$ and $\beta$ directions by $n_\alpha$ and $n_\beta$, respectively. As a representative example, we set $r_\text{o}=\SI{105}{\milli\metre}$, $r_\text{i}=\SI{35}{\milli\metre}$, $n_\alpha = 9$, and $n_\beta = 4$. 

\begin{figure}[h]
    \centering
    \includegraphics[width=0.9\columnwidth]{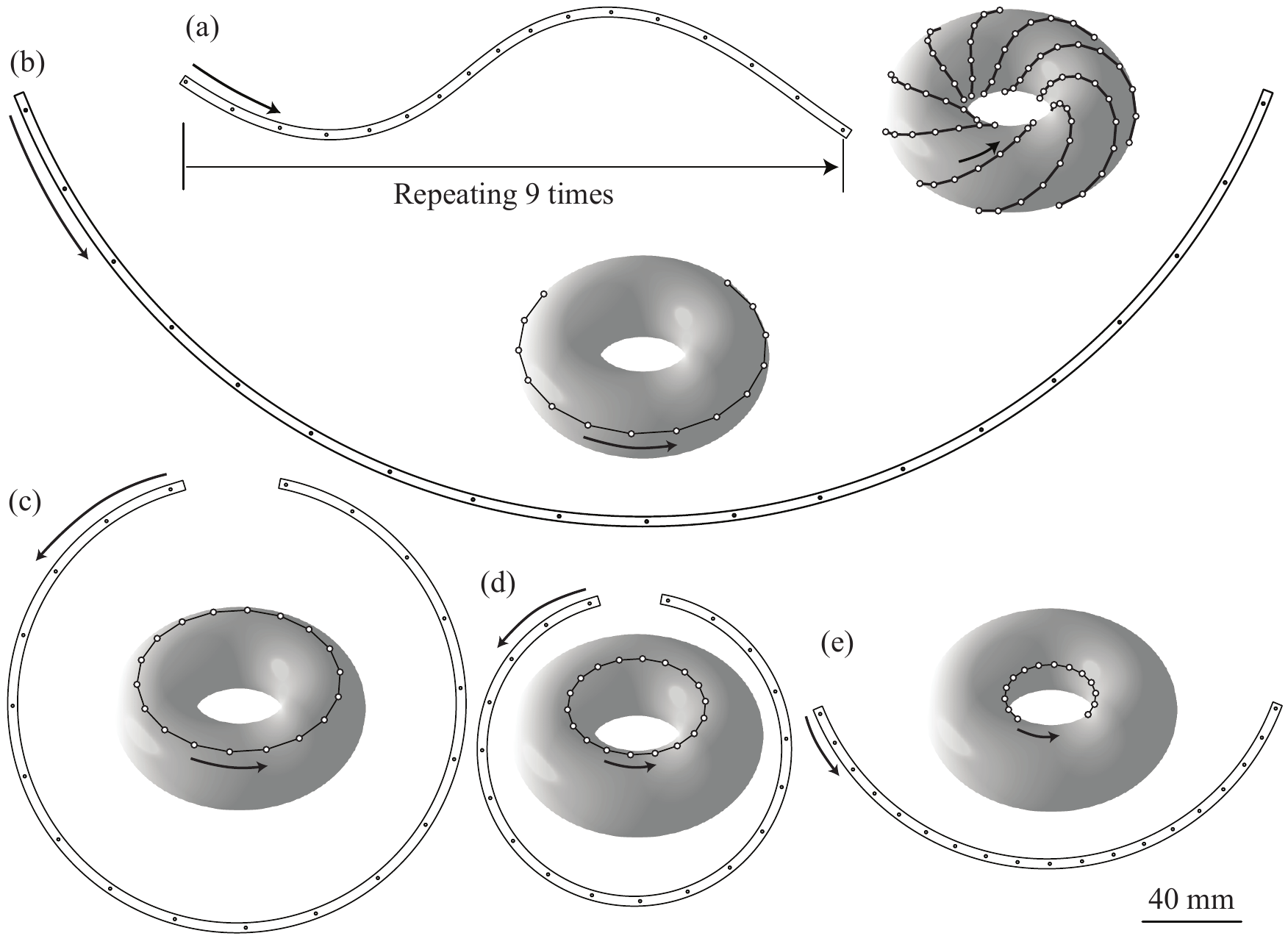}
    \caption{The ribbons comprising the toroidal weave presented in the main text: $r_\text{o}=\SI{105}{\milli\metre}$, $r_\text{i}=\SI{35}{\milli\metre}$, $n_\alpha = 9$, and $n_\beta = 4$. The weave consists of ten different ribbons; only five ribbons -- from (a) to (e) -- are presented due to symmetry. Note that the ribbons shown, all share the same scale bar ($\SI{40}{\milli\metre}$). The insets illustrate placement of ribbons.}
    \label{fig:torus_blueprint}
\end{figure}

The regular triaxial pattern in Fig.~\ref{fig:torus_topology}(a) is mapped onto the topological layout of the torus in Fig.~\ref{fig:torus_topology}(b) through Eq.~\eqref{eq:torus}, while preserving the hexagonal pattern. The topology of the weave and the target surface is presented in Fig.~\ref{fig:torus_topology}(b). To better visualize the topological layout, we color-coded the edges in both Figs.~\ref{fig:torus_topology}(a)-(b). The red and blue ribbons swirl around the torus, and the black ribbons lie on the constant-$\beta$ curves on the torus. We found that, when $n_\alpha$ and $n_\beta$ are co-prime, both the red and blue segments form single contiguous ribbons. As a result, for our specific choice of parameters ($n_\alpha = 9$, $n_\beta = 4$), there is a total of ten ribbons -- one red ribbon, one blue ribbon, and eight black ribbons -- for this specific toroidal weave. Even when $n_\alpha$ and $n_\beta$ are not co-prime, it is feasible to generate a topological layout of toroidal weaves in the same way for arbitrary values of $n_\alpha$ and $n_\beta$. For example, for the toroidal weave with $n_\alpha=10$ and $n_\beta=4$, there is a total of twelve ribbons -- two red ribbons, two blue ribbons, and eight black ribbons.

Using the protocol introduced in Sec.~\ref{subsec:inversedesign}, we computed the undeformed shape of the ribbons comprising the toroidal weave presented in Fig.~\ref{fig:torus_blueprint}; the placement of the ribbons on the weave is denoted by the black arrows on each ribbon and all the ribbons are drawn to scale. Among the ten ribbons of the weave, we only present five ribbons in Figs.~\ref{fig:torus_blueprint}(a)-(e), since the others are mirror-images due to the symmetry. Note that the ribbon shown in Fig.~\ref{fig:torus_blueprint}(a) is repeated $n_\alpha$ times. Although we designed a toroidal weave for a specific set of parameters, $r_\text{o}=\SI{105}{\milli\metre}$, $r_\text{i}=\SI{35}{\milli\metre}$, $n_\alpha = 9$, and $n_\beta = 4$, the design of toroidal weaves presented that we have introduced can be generalized to other toroidal weaves with arbitrary combinations of the parameters.

\FloatBarrier
